\documentclass[a4paper,11pt]{article}
\pdfoutput=1 

\usepackage{jheppub} 

\usepackage[T1]{fontenc} 

\usepackage{physics}
\usepackage[dvipsnames, svgnames,  x11names ]{xcolor}
\usepackage{tikz}
\usetikzlibrary{arrows,snakes,backgrounds}
\newcommand{\RNum}[1]{\uppercase\expandafter{\romannumeral #1\relax}}

\title{\boldmath Black hole interior Petz map reconstruction and
Papadodimas-Raju proposal}

\author{Niloofar Vardian

}

\affiliation{SISSA, \\via Bonomea 265, 34136 Trieste, Italy}
\affiliation{INFN, Sezione di Trieste,
\\via Valerio 2, 34127 Trieste, Italy}

\emailAdd{nvardian@sissa.it}

\abstract{We study the reconstruction of the bulk operators in AdS/CFT when the geometry contains a black hole. The black hole exterior can be mapped to the CFT via a very simple Petz map which coincides with the HKLL map reconstruction of the black hole exterior. 
For the interior modes of the bulk theory, using the definition of the Petz recovery channel in modular theory, we can find the mapping from the black hole interior to the dual boundary theory. In the case of the evaporating black hole, it is expected that the interior modes map to some operators that have support only on the bath system, the cavity that absorbs the Hawking radiation.  The most important observation that we have here is that in the case that we have a typical black hole microstate in the bulk, the CFT dual of the interior modes that we can find using the Petz recovery channel are exactly the operators that so-called "$mirror$ $operator$ " in the Papadodimas-Raju proposal. Therefore, we can interpret Papadodimas-Raju proposal as an example of the Petz map reconstruction. It may help us answer some open questions about their procedure.   
}

\begin{document} 
\maketitle
\flushbottom

\section{Introduction}


To describe our universe, we need to seek a theory of quantum gravity. The study of black hole physics can be the simplest way toward this big aim. The black hole information paradox \cite{hawking1974black, hawking1975particle} is one of the important questions of physics in the last few decades \cite{almheiri2013black,harlow2013quantum, mathur2014flaw, almheiri2013apologia, van2014evaporating,marolf2013gauge, verlinde2013black, verlinde2013passing,verlinde2013black}. It is related to the smoothness of the horizon of the black hole.
Among all the work related to it, I will focus on  the Papadodimas-Raju proposal that the authors could even construct the interior operator on the boundary side \cite{papadodimas2013infalling, Papadodimas:2013jku,Papadodimas:2013wnh, Papadodimas:2015jra} , they called it $mirror$ $operator$
\footnote{The mirror operator in the Papadodimas-Raju proposal is different from the mirror operators that are defined in the appendix.}.

In parallel, we have the island conjecture related to the information paradox and address very abstractly that we can construct the modes in the island from the Hawking radiation by mapping the so-called Petz map \cite{cotler2019entanglement}.
The Petz map has its origin in quantum information theory \cite{petz1986sufficient, petz1988sufficiency}. Recently, it has been found that the best way to understand the semi-classical limit of AdS/CFT is in the language of quantum error correction codes \cite{dong2016reconstruction}. The error-correcting codes then are the isometries from the bulk Hilbert space to the dual boundary theory.  

In the case of subregion duality, at large N limit, following the JLMS argument \cite{jafferis2016relative}, there must be a recovery channel that maps the operator in the entanglement wedge $ a = \mathcal{E}_A$ to the given region of the boundary $ A$. 
In \cite{cotler2019entanglement}, authors could find the explicit expression of the recovery channel using the global HKLL map as a global isometry that embeds the entire bulk to the entire boundary which is known as the Petz map. However, it was very recently that the explicit calculation using the Petz map has been done \cite{Bahiru:2022ukn}.

When the geometry contains a black hole, because of the lack of such an isometry we can not follow the discussion in \cite{cotler2019entanglement} and write the explicit form of the mapping. 
In this paper, instead of following that work to write the quantum channel by taking trace over the complementary region, we use the definition of the Petz map in modular theory. It is good to note that the Petz recovery channel has it is origin in modular theory, roughly speaking at the age when the theory of quantum computation and information was born.
We find the Petz reconstruction of the interior modes and we reach the same result as the Papadodimas-Raju proposal. 

In this paper, in  Sec. \ref{Sec.Petz}, we review the Petz recovery channel. In Sec. \ref{sec2}, we will describe some aspects of black hole physics in AdS, and then, in Sec. \ref{petz,bh} we will discuss in detail how we can use the notion of Petz map in modular theory to reconstruct the interior modes in the geometries contain a black hole. We start with the two-sided eternal black hole in AdS and then we will generalize the discussion to the one-sided one.
Finally, in Sec. \ref{PR} we will discuss the connection with the Papadodimas-Raju proposal for the reconstruction of the black hole interior in AdS/CFT.

\section{Petz map}\label{Sec.Petz}

In this section, we introduce the Petz recovery channel and in particular we review its original definition in modular theory.

\subsection{Universal recovery channel}

In quantum information theory, the evolution of states in the Schrodinger picture is modeled by a quantum channel $ \mathcal{E}: \mathcal{S}(\mathcal{H}_A) \rightarrow \mathcal{S}(\mathcal{H}_B) $, where $\mathcal{S}(\mathcal{H}) $ is the set of all density matrices on the Hilbert space $ \mathcal{H}$.
The channel $ \mathcal{E}$ is reversible if there exists another quantum channel $ \mathcal{R}$, called the \emph{recovery channel} 
such that 
$ \mathcal{R} \circ \mathcal{E} (\rho)=\rho$  for all $\rho \in \mathcal{S}(\mathcal{H}_A)$.
The reversibility of $ \mathcal{E}$ is related to the quantum relative entropy of states under the action of $ \mathcal{E}$. The relative entropy between two states is a measure of distinguishability of them  which is defined as
\cite{umegaki1962conditional}
\begin{equation}\label{26}
   S(\rho | \sigma) = \tr\big(\rho \log \rho - \rho \log \sigma \big). 
\end{equation}
It is a non-increasing function under the action of any quantum channel, $i.e.$
\begin{equation}\label{3}
    S(\rho | \sigma)\geq S ( \mathcal{E}(\rho)| \mathcal{E}(\sigma))
\end{equation}
that is known as \emph{data processing inequality} \cite{lieb1973fundamental,lindblad1975completely}.
For a given channel $ \mathcal{E}$, we have the exact correctability just in case the equality in (\ref{3}) holds
\cite{jenvcova2006sufficiency, jenvcova2006sufficiency2, petz1986sufficient, petz1988sufficiency}.
In such a case, the recovery channel is given in terms of the dual channel $ \mathcal{E}^*$ and
once fixed full rank density matrix $ \rho \in \mathcal{S}(\mathcal{H}_A)$ 
\begin{equation}\label{4}
    \mathcal{P}_{\rho,\mathcal{E}}(.)=\rho ^{1/2} \mathcal{E}^* \big(\mathcal{E}(\rho)^{-1/2} (.)\mathcal{E}(\rho)^{-1/2}\big)\rho^{1/2}
\end{equation}
known as the \emph{Petz recovery channel}
\cite{petz2003monotonicity}.
In \eqref{4}, $ \mathcal{E}^*$ is the (Hilbert-Schmidt) dual channel  defined as the solution to 
\begin{equation}\label{dual}
   \tr \big( \rho~ \mathcal{E}^*(O)\big)= \tr \big( \mathcal{E}(\rho)~O\big) 
\end{equation}
for all $ \rho \in \mathcal{S}(\mathcal{H}_A)$ and $ O\in \mathcal{L}(\mathcal{H}_B)$. The set of all bounded operators act on $ \mathcal{H}$ is denoted by  $  \mathcal{L}(\mathcal{H})$.
One can use the Heisenberg picture and map the operators by using the dual of the recovery channel
$ \mathcal{R}^*: \mathcal{L}(\mathcal{H}_B) \rightarrow \mathcal{L}(\mathcal{H}_A) $,
which for the quantum channel in \eqref{4} is given as 
\begin{equation}
    \mathcal{P}_{\rho,\mathcal{E}} ^* (.)
    =\mathcal{E}(\rho )^{-1/2}
     \mathcal{E}  \big(\rho ^{1/2} (.) \rho ^{1/2}\big)\mathcal{E}(\rho)^{-1/2}
\end{equation}
called \emph{Petz map}.

\subsection{Petz map and modular theory}

In order to study the recovery of the information in quantum field theories, it would be really helpful to have an alternative description for the Petz recovery channel in the 
context of the Tomita-Takesaki theory (A brief review can be found in the Appendix \ref{app}).
Luckily, the Petz recovery channel indeed has its origin in the study of operator
algebras \cite{ohya2004quantum,petz1986sufficient, petz1988sufficiency}. It has also been studied in some recent works \cite{furuya2020real, faulkner2022approximate,faulkner2022approximate2}.
We will now review the definition of the Petz map in the algebraic approach. All the discussion below is in the Heisenberg picture and it is almost based on \cite{Furuya:2020tzv}. 

Consider two Type \RNum{1} von Neumann algebras 
in their standard forms:
$(\mathcal{A}, \mathcal{H}_A, J_A, \mathcal{P}_{\mathcal{A}} )$ and 
$ (\mathcal{B},   \mathcal{H}_B, J_B, \mathcal{P}_{\mathcal{B}} )$.
Assume two faithful states $ \rho_A$ and $ \rho _B$ respectively on $ \mathcal{A}$ and $ \mathcal{B}$ that we will refer to their unique vector representations  by 
$ |\rho  _A ^{1/2}\rangle \in \mathcal{P}_A$ 
and $ |\rho_B ^{1/2} \rangle \in \mathcal{P}_B$.
The corresponding GNS Hilbert spaces of the algebras over the states $ |\rho  _A ^{1/2}\rangle $ and $ |\rho  _B ^{1/2}\rangle $ are denoted by $ \mathcal{H}_A$ and  $ \mathcal{H}_B$.

Let us consider  a linear superoperator $ \mathcal{T}: \mathcal{A} \rightarrow \mathcal{B}$ and denote its corresponding operator between the corresponding  GNS Hilbert spaces by $ T : \mathcal{H}_A \rightarrow \mathcal{H}_B$.
One can define a dual of it $ \mathcal{T}^*_\rho: \mathcal{B} \rightarrow \mathcal{A}$  as a solution to 
\begin{equation}\label{5}
    \langle b \ket{ \mathcal{T}(a)} _ {\rho _B} =  \langle b \ket{Ta} _ {\rho _B} =  \langle T^\dagger b \ket{a} _ {\rho _A} =  \langle \mathcal{T}^*_{\rho}(b) \ket{a} _ {\rho _A}
\end{equation}
in the GNS Hilbert space \eqref{27} for all $ a \in \mathcal{A}$ and $ b \in \mathcal{B}$ .
In the case of matrix algebra, the definition (\ref{5}) can be rewritten as
\begin{equation}\label{22}
    \tr (\rho_B b^\dagger \mathcal{T}(a)) = \tr ( \rho_A \mathcal{T}^* _\rho (b^\dagger) a).
\end{equation}
We note it here that if we replace both $ \rho _A$ and $ \rho _B$ with the identity operators (unnormalized maximally mixed states), the dual map $ \mathcal{T}^*$ we will get is the usual dual map in the quantum information theory defined in \eqref{dual}.
One can find  $\mathcal{T}^*_\rho$ in \eqref{22} in terms of $ \mathcal{T}^*$ as 
\begin{equation}
     \mathcal{T}^*_\rho (b) = \rho_A ^{-1} \mathcal{T}^* ( \rho _ B b).
\end{equation}
On the other hand, the GNS Hilbert space $\mathcal{H}_A$ can be created by acting with the commutant of the algebra $ \mathcal{H}_A = \{ \mathcal{A}' ~|\rho_A ^{1/2} \rangle\}$ and the same for another algebra $ \mathcal{H}_B =\{ \mathcal{B}' ~|\rho_B ^{1/2} \rangle\} $. 
Therefore, given $ T$ between the GNS Hilbert spaces, we can in principle associate to it  one superoperator between the commutants $ \mathcal{T}'_\rho: \mathcal{B}' \rightarrow \mathcal{A}'$ which defined as 
\begin{equation}\label{6}
    \langle b' \ket{ \mathcal{T}(a)} _ {\rho _B} =    \langle \mathcal{T}'_{\rho}(b') \ket{a} _ {\rho _A}
\end{equation}
for all $ a \in \mathcal{A}$ and $ b' \in \mathcal{B}'$ that is called $\rho$-$dual$ of the superoperator $ \mathcal{T}$.
Then, one can use the modular conjugations in both Hilbert spaces to define a superoperator between the original algebras $\mathcal{T}^P_{\rho}:\mathcal{B} \rightarrow \mathcal{A}$ as 
\begin{equation}\label{7}
    \mathcal{T}^P_{\rho}(.) = \mathcal{J}_A \circ \mathcal{T}'_{\rho} \circ \mathcal{J}_B (.) = J_A ~\mathcal{T}'_{\rho} ~\big(J_B~ (.)~J_B \big) ~J_A
\end{equation}
which is exactly the Petz dual map we are interested in that for the Type \RNum{1} von Neumann algebra, one can explicitly find its form as
\begin{equation}\label{23}
    \mathcal{T} ^P_\rho (.)
    =\rho _A ^{-1/2}
     \mathcal{T} ~ \big(\rho_B ^{1/2}~ (.) ~\rho_B ^{1/2}\big)\rho _A ^{-1/2}.
\end{equation}
The Petz dual map can also be realized as the solution to the relation \eqref{5}
 with respect to the KMS inner product. If the state $ \ket{\Psi}$ is cyclic and separating for a von Neumann algebra $ \mathcal{A}$, the KMS inner product on $ \mathcal{A}$ is defined as 
\begin{equation}\label{45}
    \langle a_1 \ket{ a_2}_{\psi, \text{KMS}} = \langle \mathcal{J}_\psi(a_1^\dagger)\ket{a_2}_\psi
    = \bra{\Psi} a_1^\dagger ~\Delta_\psi^{1/2} a_2 \ket{\Psi}.
\end{equation}
while the last expression can be found using \eqref{24}.
In the case of matrix algebra, it is reduced to 
$ \langle a_1 \ket{a_2}_{\rho} = \tr ( \rho ^{1/2} a_1^\dagger \rho ^{1/2} a_2)$.

While the definition \eqref{23} is only for Type \RNum{1} von Neumann algebras, the one in \eqref{7} can be generalized to the mapping between general von Neumann algebras. In particular, it is helpful for high-energy physics applications where the von Neumann algebras under consideration are of Type \RNum{3}$_1$.

To summarize, consider  $  (\mathcal{A}, \mathcal{H}_A, J_A, \mathcal{P}_{\mathcal{A}} )$, $ (\mathcal{B}, \mathcal{H}_B, J_B, \mathcal{P}_{\mathcal{B}} )$ and let $ \mathcal{T}: \mathcal{A}\rightarrow \mathcal{B}$ be a unital completely positive map between the algebras.
 One can choose an arbitrary state $ \rho_B \in \mathcal{S}(\mathcal{B})$ and if both $\rho _B$ and $ \rho _A = \mathcal{T}^*(\rho_B)$ are faithful states on the corresponding von Neumann algebras, construct the Hilbert space representation of the algebras over them. Then,
if for all $ \rho , \sigma \in \mathcal{S}(\mathcal{B})$ we have 
\begin{equation}
    S(\rho | \sigma) = S( \mathcal{T}^*(\rho) | \mathcal{T}^*(\sigma)),
\end{equation}
there exists a unital completely positive map $ \Tilde{\mathcal{T}}: \mathcal{B}\rightarrow \mathcal{A}$ that $ \mathcal{T} \circ \Tilde{\mathcal{T}}$ acts as an identity operator on $ \mathcal{H}_B$. The $\Tilde{\mathcal{T}}$ is nothing but the Petz dual map given in (\ref{7}) 
 and can be also shown that  
$  (\Tilde{\mathcal{T}})^* = \mathcal{P} _{\rho_{B}, \mathcal{T}^*} $ in \eqref{4}.
In \cite{faulkner2022approximate,faulkner2022approximate2}, one can find the discussion for generalization of the Petz dual map in cases where the states are not faithful.

\subsection{Approximate recoverability}\label{approx}

The equality of relative entropy is a necessary and sufficient condition for exact recoverability. In the case of approximate quantum error correction, the quality of recovery is controlled by the behavior of the relative entropy under the action of the quantum channel. At the heart of this result is a strengthened version of the monotonicity of the relative entropy which undergoes a slight change in relative entropy through the channel provides the approximate recoverability of the states. 

For a given channel $ \mathcal{E}$, Junge et al \cite{junge2018universal} found an expression for the recovery channel $ \mathcal{R}_{\rho, \mathcal{E}} $, which is closely related to Petz recovery channel and it is also universal. In terms of $ \mathcal{E}$ and an arbitrary full rank density matrix $ \rho$, it is given by 
\begin{equation}\label{29}
   \mathcal{R}_{\rho, \mathcal{E}}(.) = \int _{-\infty}^{ \infty} dt~ p(t)~
   \rho^{-it}~ \mathcal{P}_{\rho, \mathcal{E}} \big( \mathcal{E}(\rho)^{it}~ (.) \mathcal{E}(\rho)^{-it}\big) ~ \rho^{it}
\end{equation}
while $ p(t) = \pi /( \cosh (2\pi t) +1) $ and 
$\mathcal{P}_{\rho, \mathcal{E}}$ is the Petz recovery channel given in \eqref{4}. Moreover, they gave a lower bound on the difference between the relative entropy in terms of the fidelity between the original state and the recovered one as
\begin{equation}
    S(\rho | \sigma) - S( \mathcal{E}(\rho)| \mathcal{E}(\sigma)) \geq
    -2 \log F( \rho , \mathcal{R}_{\sigma, \mathcal{E}} \circ \mathcal{E} (\rho)).
\end{equation}

The result above has been found in the context of the quantum information theory $i.e.$, Type \RNum{1} von Neumann algebra. While one should go beyond it 
in the case of QFTs and gravity. 
In \cite{ faulkner2022approximate,faulkner2022approximate2}, authors generalized the previous result to quantum channels between general von Neumann algebras in the context of modular theory.  
In the Heisenberg picture, consider $ \mathcal{T} : \mathcal{A}\rightarrow \mathcal{B}$ be a unital, normal and two-positive map between von Neumann algebras. 

One can associate a dual Petz map  $  \mathcal{T} ^P_\psi :  \mathcal{B} \rightarrow \mathcal{A}$ \eqref{7}. 
They found the recovery map in the form of 
\begin{equation}\label{28}
     \alpha (.) = \int _{-\infty}^{\infty} dt~ p(t)~ \alpha^t_{\psi, \mathcal{T}} (.)  
\end{equation}
while  
\begin{equation}
    \alpha^t_{\psi, \mathcal{T}} (.) = \zeta ^t _{\psi, \mathcal{A}}\circ 
    \mathcal{T}^P_{\psi}
    \circ\zeta ^{-t} _{\psi, \mathcal{B}}  (.)
\end{equation}
is called the \emph{rotated Petz map}, and $ \zeta ^t _{\psi, \mathcal{A}}$ is the modular flow for $ \ket{\Psi_A}, \mathcal{A}$ is define as
\begin{equation}
     \zeta ^t _{\psi, \mathcal{A}} (a) =Ad~ \Delta^{it} _{\psi, \mathcal{A}}~ (a) =  \Delta^{it} _{\psi, \mathcal{A}}~ (a) ~\Delta^{-it} _{\psi, \mathcal{A}} \qquad\qquad \forall a \in \mathcal{A}.
\end{equation}
In the case of finite-dimensional Type \RNum{1} factor, the recovery map in \eqref{28} reduced to the dual of the recovery channel in \eqref{29}.

\section{Black holes in AdS}\label{sec2}

In this section, We briefly review the geometry of the eternal two-sided AdS black holes, quantizing the free field theory on this background which is needed for studying the bulk reconstruction in AdS/CFT at strict large N limit. 
We also discuss the operator algebra of observable in the case that an eternal black hole in the bulk is dual to the two CFTs in the termo-field-double (TFD) state, and in the end, talk about the typical one-sided black holes in AdS.

\subsection{AdS eternal black holes}

\begin{figure}[h]
    \centering
    
    \begin{tikzpicture}
\shadedraw[left color= lightgray,right color=white, draw=darkgray](0,0)--(3,3)--(3,-3)--cycle;
\shadedraw[right color= lightgray,left color=white, draw=darkgray](0,0)--(-3,3)--(-3,-3)--cycle;

\draw[Crimson, thick, snake=snake, segment amplitude=.4mm, segment length=2mm] (-3,3)--(3,3);
\draw[Crimson, thick, snake=snake, segment amplitude=.4mm, segment length=2mm]  (-3,-3)--(3,-3);

\node at (2.3,-1.5) {${\RNum{1}}$};
\node at (0.3,2) {${\RNum{2}}$};
\node at (-2.3,1.5) {${\RNum{3}}$};
\node at (-0.3,-2) {${\RNum{4}}$};

\draw[DarkGoldenrod2, very thick] (-3,0) .. controls (-1.5,0.2) .. (0,0);
\draw[DarkGoldenrod2, very thick] (0,0) .. controls (1.5,-0.2) .. (3,0);
\node at (-2.2,-0.4) {$\Sigma_{l}$};
\node at (2.2,0.4) {$\Sigma_r$};
\node at (3.4,0.2) {$\Sigma$};
\end{tikzpicture}

\caption{The Penrose diagram of the eternal two-sided black hole.}
\label{eternalBH}
   
\end{figure}
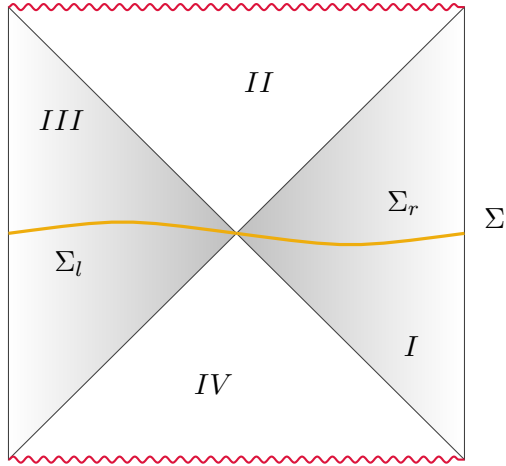

There is a unique spherically symmetric solution of the Einstein equation with negative curvature known as AdS-Schwarzschild geometry. Its metric in $(d+1)$-dimension ($d\geq3$) is given by
\begin{equation}\label{10}
    ds^2= - f(r) dt^2 + \frac{dr^2}{f(r)} + r^2 d\Omega^2_{d-1}.
\end{equation}
In the AdS unit, we have
\begin{equation}
    f(r) = 1+ r^2 - \frac{\alpha}{r^{d-2}}
\end{equation}
where $\alpha$ is a parameter proportional to the ADM mass $M$.
Like in flat space, one can maximally extend the solution by introducing AdS-Kruskal coordinates
\begin{equation}
    \begin{split}
        U\equiv & -e^{2\pi(r_*-t)/ \beta}
        \\
        V\equiv & \; e^{2\pi(r_*+t) /\beta}
    \end{split}
\end{equation}
where here, $ r_*$  is the tortoise coordinate defined as $ \frac{dr_*}{dr}= f^{-1}(r)$. 
The metric in the new coordinates can be written as 
\begin{equation}\label{11}
     ds^2= (\frac{\beta}{2\pi})^2\frac{ f(r)}{UV} dU dV + r^2 d\Omega^2_{d-1}.
\end{equation}
The metric is originally defined in the region $U<0$, $ V>0$ corresponding to outside the horizon.
By extending the geometry in a maximal way when we assume that there is no matter anywhere, one can describe eternal black holes in AdS spacetime (Fig. \ref{eternalBH}).
In all regions, one can introduce Schwarzschild coordinates. Their relation with Kruskal coordinates is given as \cite{Papadodimas:2015jra} :

\begin{center}
\begin{tabular}{ c|c|c}
regions & Kruskal coordinates & relationship with the AdS-Schwarzschild coordinates\\
\hline
\RNum{1} & $ U< 0 , V> 0$ & $ U= -e^{\frac{2\pi}{\beta}(r_*-t)}, V= e^{\frac{2\pi}{\beta}(r_*+t)}$\\
\RNum{2} & $ U> 0 , V> 0$ & $ U= e^{\frac{2\pi}{\beta}(r_*-t)}, V= e^{\frac{2\pi}{\beta}(r_*+t)}$\\
\RNum{3} & $ U> 0 , V< 0$ & $ U= e^{\frac{2\pi}{\beta}(r_*-t)}, V= -e^{\frac{2\pi}{\beta}(r_*+t)}$\\
\RNum{4} & $ U< 0 , V< 0$ & $ U= -e^{\frac{2\pi}{\beta}(r_*-t)}, V= -e^{\frac{2\pi}{\beta}(r_*+t)}$

\end{tabular}
\end{center}

Regions \RNum{1} and \RNum{3} are two asymptotically AdS regions corresponding to black hole exteriors that for each of them, another one is behind the horizon.
In the U-V plane, surfaces of constant $r_*$ are hyperboloids that always stay within a single region. On the other hand, the surfaces of constant $t$ are simply straight lines running through the origin which means we can think of time translations as rotations of the Kruskal diagram about the bifurcation point.
Although, we should keep in mind that a line can not be rotated past the horizon by a finite rotation.
Moreover, since there is no global timelike isometry, the entire geometry is time-dependent.

It is good to note that the natural choice for the vacuum of the bulk effective theory on the AdS eternal black hole is the Hartle-Hawking (HH) state $ \ket{HH}$. 
It has been conjectured by Maldacena 
\cite{maldacena2003eternal} that the AdS eternal black hole has a holographic description in terms of two copies of an identical CFT in the TFD state 
\begin{equation}\label{21}
    \ket{\Psi_{TFD}} = \frac{1}{\sqrt{Z_\beta}} \sum _i e^{-\beta E_i /2} \ket{i^*}_L \ket{i}_R.
\end{equation}
where $\beta^{-1}$ is the Hawking temperature of the black hole.
Therefore, one can describe each holographic CFT dual to the eternal black hole with the thermal density matrix
$ \rho_{th} = \frac{1}{Z_\beta} e^{-\beta H}$, where $ Z_\beta$ is the partition function of the CFT at the temperature $\beta^{-1}$.
Here, the states $ \ket{i}$ are the energy eigenstates of one single CFT and 
$ \ket{i^*}_{L(R)} = \Theta\ket{i}_{R(L)}$. 
The  $\Theta$ is an anti-unitary operator that reverses the time direction after exchanging the CFTs.
Having the states $ \ket{i^*}$ for the left CFT comes from the point that the left CFT is glued to the region \RNum{3} with a flip in the AdS-Schwarzschild time coordinate.  In other words, the time in left and right CFTs are identified as $ t_R= t$ and $ t_L = -t$, respectively.
The isometry of the entire bulk generated by $ t \rightarrow t+T$ also corresponds to the identity
\begin{equation}\
     e^{i \hat{H}}{ \ket{\Psi_{TFD}}} = \ket{\Psi_{TFD}}, \qquad\qquad \hat{H}= H_R -H_L.
\end{equation}
The time translation on the entire geometry is generated by a Killing vector field denoted by $V$. It is  future-directed timelike on the right exterior and  past-directed timelike on the left exterior. The conserved charge associated to the global time translation is 
\begin{equation}\label{30}
    \hat{h} = \int _\Sigma d \Sigma^\mu~ V^\nu T_{\mu\nu}
\end{equation}
where $ \Sigma$ is a Cauchy hypersurface and $ T_{\mu \nu}$ is the energy-momentum tensor of the bulk fields.
The boundary dual of the operator $\hat{h}$ is 
\begin{equation}\label{33}
    \hat{h} = \beta \hat{H}.
\end{equation}
It comes from the fact that the vector field $V$ on the right boundary reduces to $ \beta \partial_t$ and to $ - \beta \partial_t$ on the left one. We will discuss these quantity's interpretation in the algebraic context later.

\subsection{Scalar field quantization in AdS eternal black hole background}

Consider a free scalar field propagating on a curved spacetime background. The equation of motion is the Klein-Gordon (KG) equation, $ (\Box-m^2)\phi =0$. The field then has a Heisenberg picture expression as
\begin{equation}
    \phi (x) = \sum _n~ f_n(x)~ a_n +  f^*_n(x)~ a_n^\dagger,
\end{equation}
where $ f_n$ are the classical solutions of the KG equation in the given background that should be normalized with respect to the KG norm. We should also impose normalizable boundary conditions at infinity.
To each mode $f_n$, we associate the annihilation and creation operators  $a_n$, $a_n^\dagger$  with normalized commutation relation. The Hilbert space of QFT at every Cauchy slice of the entire background of interest can be constructed as a Fock space using these ladder operators.

On the other hand, we can always decompose a Cauchy slice $\Sigma$ into the smaller slices $\Sigma_r$ such that their union covers the entire Cauchy surface, $ \Sigma= \cup_r\Sigma_r$. 
We can find an alternative expression for the field in the domain of dependence of each subregion denoted by $\mathcal{D}(\Sigma_r)$, by solving the equation of motion on the coordinate system that covers only $ \mathcal{D}(\Sigma_r)$
\begin{equation}
     \phi (x_r) = \sum _\alpha~ f_{r, \alpha}(x_r)~ a_{r, \alpha} +  f^*_{r, \alpha}(x_r)~ a_{r, \alpha}^\dagger.
\end{equation}
The new operators $a_{r, \alpha} $ have support only on $\Sigma_r$.
Here, we can use the Bogoliubov transformation and write the mode functions on the entire Cauchy slice as  linear combinations of $ a_{r, \alpha} $. Hence, we can in principle expand the field $ \phi$ at each point in terms of $a_{r, \alpha} $.

Now, let us quantize a scalar field on the eternal AdS black hole background. In principle, we can do it by solving the equation of motion in Kruskal coordinates. However, there is another possibility when we take the Cauchy slice $\Sigma$ such that it passes through the bifurcation point. We can consider $\Sigma$ as the union of two smaller slices as
$ \Sigma= \Sigma_r \cup \Sigma_l$ (Fig. \ref{eternalBH}). Then, if we consider just region \RNum{1}(\RNum{3}), $\Sigma_{r} (\Sigma_{l})$ itself is a complete Cauchy slice on that. We also know coordinate systems that cover regions \RNum{1} and \RNum{3} which are nothing but two copies of AdS-Schwarzschild coordinates. Therefore, we can first start by region \RNum{1} and solve the KG equation outside the horizon in AdS-Schwarzschild coordinates 
(\ref{10}).
One can find its solutions as
$f_{\omega,m}$ which are the modes labeled by quantum numbers $ \omega,m$. To each of them, we associate a couple of creation and annihilation operators with normalized commutation relation, denoted by $ a_{\omega,m}$. Therefore, one can express the fields lies in region \RNum{1} as
\begin{equation}\label{8}
  \phi_{\RNum{1}} (x) =  \sum _m \int_0^{\infty} \frac{d\omega}{2\pi} \frac{1}{\sqrt{2\omega}} 
 ~ \big(f_{\omega,m}(x)a_{\omega,m} + f^*_{\omega,m}(x)a^\dagger_{\omega,m} \big).
\end{equation}
We can follow the same analysis in region \RNum{3} and find another set of operators $ \Tilde{a}_{\omega,m}$ with the same algebra as $ a_{\omega,m}$ while they commute with all $a_{\omega,m}$. One can write the fields in region \RNum{3} like region \RNum{1} as
\begin{equation}\label{9}
  \phi_{\RNum{3}} (x) =  \sum _m \int_0^{\infty} \frac{d\omega}{2\pi} \frac{1}{\sqrt{2\omega}}
  ~\big(\Tilde{f}_{\omega,m}(x) \Tilde{a}_{\omega,m} + \Tilde{f}^*_{\omega,m}(x)\Tilde{a}^\dagger_{\omega,m}\big).
\end{equation}
Therefore, we have the expansion of the field in the entire Cauchy slice $\Sigma$ and so, it is straightforward to find the expression for fields in region \RNum{2} and \RNum{4} by evolving them with respect to the total Hamiltonian.

As it is mentioned, the vacuum of the quantum field in the eternal black hole background is an analog of the HH state 
corresponding to the black hole temperature $ T = 1/\beta$
which is defined to satisfy 
\begin{equation}
    \big(a_{\omega,m} - e^{-\beta \omega 2} a^\dagger _{\omega, m}\big) \ket{HH} =0,
\end{equation}
and characterized by thermal occupation levels for both modes $a_{\omega,k}$ and $ \Tilde{a}_{\omega,k}$
\begin{equation}\label{15}
    \begin{split}
        \langle a_{\omega,m}  a^\dagger_{\omega',m'}\rangle_{HH} = \frac{e^{\beta\omega}}{e^{\beta\omega}-1} \delta(\omega-\omega')\delta_{m,m'}
        \\
        \langle a^\dagger_{\omega,m}  a_{\omega',m'}\rangle_{HH} = \frac{1}{e^{\beta\omega}-1} \delta(\omega-\omega')\delta_{m,m'}
    \end{split}
\end{equation}
and the same for the modes $ \Tilde{a}_{\omega,m}$ \cite{Papadodimas:2012aq}.
Using this standard formalism of quantization of the field theory on the curved spacetime background, one can also describe the Hilbert space of the theory as the Fock space built on the HH vacuum, denoted as $ \mathcal{H}_{BH}^{(Fock)}$.

\subsection{Operator algebra of observables}

Consider two copies of the boundary CFT in the TFD state dual to a two-sided eternal black hole in AdS. 
As explained in the previous section, small perturbations around the black hole background can be described by quantizing the QFT in a curved spacetime background.
The algebra of low-energy effective field theory in the left and right exteriors is denoted by $ \mathcal{A}_{l,0}$ and $ \mathcal{A}_{r,0}$ respectively.
In the large N limit, the algebras of observables outside the horizon of the black hole, i.e. $ \mathcal{A}_{l,0}$ and $ \mathcal{A}_{r,0}$, are of Type \RNum{3}$_1$ von Neumann algebra since the algebra of observables of a local region in QFT must be of this Type. As these two spacetime regions are spacelike separated, the corresponding operator algebras are each other's commutants.

One can split $ \hat{h}$ in \eqref{30} as a difference between right and left operators as 
\begin{equation}\label{31}
    \hat{h} = h_r - h_l
\end{equation}
while 
\begin{equation}
    h_r= \int _{\Sigma_r} d \Sigma^\mu~ V^\nu T_{\mu\nu},\qquad\qquad  h_l= \int _{\Sigma_l} d \Sigma^\mu~ V^\nu T_{\mu\nu}.
\end{equation}
To obtain this splitting we should choose the Cauchy hypersurface $\Sigma$ to pass through the bifurcate horizon, i.e. $ \Sigma = \Sigma_r \cup \Sigma_l$. 
The question that can arise at this point is whether the operators $ h_l$ and $ h_r$ belong to the operator algebras $ \mathcal{A}_{l,0}$ and $ \mathcal{A}_{r,0}$ or not? 

As explained in \cite{witten2022gravity} besides this formal splitting, 
due to the ultraviolet divergences near the horizon, $i.e.$
\begin{equation}
    |h_{r(l)}\ket{HH}|^2 = \bra{HH} h_{r(l)}^2 \ket{HH} = \infty,
\end{equation}
the operators $ h_r, ~h_l$ do not make sense as an operator on the bulk Hilbert space $ \mathcal{H}_{BH}^{(Fock)}$.
There is another way of answering this question: in the Tomita-Takesaki theory, the operator $ \hat{h}$ is related to the modular Hamiltonian of the HH state for the algebra $\mathcal{A}_{r,0}$
\begin{equation}
    \Delta = e^{- \hat{h}}
\end{equation}
A modular Hamiltonian of a Type \RNum{3}$_1$ algebra never has a splitting as in \eqref{31} and so the operators $ h_r, ~h_l$ are not well-defined and so they do not belong to the operator algebras $\mathcal{A}_{l,0}$ and $\mathcal{A}_{r,0}$ at strict large N limit.

The operators dual to the low energy effective field theory are the subtracted single trace operators of the boundary theory which have Gaussian correlation functions in the large N limit and we denote them here as $ \mathcal{A}_{L,0}$ and $ \mathcal{A}_{R,0}$ for the left and right CFTs. 
Therefore, the commutator of the single trace operators in the large N limit is $c-$number. 
Since the operators $ h_l$ and $h_r$ are not part of the bulk operator algebras, the gauge theory Hamiltonian of the boundary theories, $ H_L$ and $H_R$ must not be part of the algebras $ \mathcal{A}_{L,0}$ and $ \mathcal{A}_{R,0}$ as well.

Above the Hawking-Page temperature that the two CFTs in the TFD state are dual to the two-sided eternal black hole in the bulk, $ H_R$ and $ H_L$ have the thermal expectation value and connected two-point function of order $ N^2$. 
Although the operators $H_R$ and $H_L$ do not have a large $N$ limit, their difference does have as its bulk dual $ \hat{h}$ \eqref{33}. The modular operator of the TFD state of the boundary for the algebra $\mathcal{A}_{R,0}$ is 
$ \Delta = e^{- \beta \hat{H}}$. To obtain the modular operator, one can also start with finite $N$ where the algebra of observables on both sides are of Type \RNum{1} von Neumann algebras. The left and right Hamiltonians can be written in terms of the usual Hamiltonian of a single copy of the system $H$ as $ H_L = H \otimes I$ and $ H_R = I \otimes H$. In the case of Type \RNum{1} algebra, each system individually can be described by a reduced density matrix which here for the TFD state, are obtained to be the thermal density matrices 
$ \rho_L = \rho_R = e^{ - \beta H} / Z$. 
Then, one can find the modular operator of the TFD state by using the relation \eqref{34} as 
\begin{equation}
    \Delta = \rho_L^{-1} \otimes \rho_R = e^{ - \beta H_R + \beta H_L} = e^{-\beta \hat{H}}
\end{equation}
which is also valid at large $N$ limit.
By subtracting the expectation value of the Hamiltonian, one can define the operator 
$ H_R' = H_R - \langle H_R \rangle$
and the same for the left side. We have 
$ \langle H_R'^2 \rangle \sim N^2  $ and thus $H_R'$ does not have a large $N$ limit. 
By dividing it by $N$, we can introduce an operator 
\begin{equation}
    U = \frac{1}{N} H_R'
\end{equation}
with Gaussian correlation function at large $N$ limit, the same as any other single trace operators.
Therefore, $U$ has a large $N$ limit, but at this limit, it is central as 
\begin{equation}
    [ U, O]= \frac{1}{N} [H_R , O] = - \frac{i}{N} \frac{\partial O}{ \partial t}
\end{equation}
and at $ N=\infty$, it commutes with all the rest of the single trace operators.

As explained, $H_R$ and thus $ U$ is not part of the algebra $ \mathcal{A}_{R,0}$. Therefore, we can also define $ \mathcal{A}_{R,0}$ to consist of only the single trace operators that have non-zero commutators.
In other words, the operator algebra of low-energy excitations around the black hole background is dual to the single trace operators of the boundary  with a nontrivial commutator,
they describe a \emph{generalized free field} (GFF) over the thermal state of the CFT. 
Using the AdS/CFT argument, we can identify the operator algebra of the bulk and boundary as 
\begin{equation}
     \mathcal{A}_{r,0} =\mathcal{A}_{R,0},\qquad\qquad  \mathcal{A}_{l,0} =\mathcal{A}_{L,0}
\end{equation}
which by itself requires that $ \mathcal{A}_{L,0}$ and $ \mathcal{A}_{R,0}$ be of Type \RNum{3}$_1$ as well \cite{witten2022gravity}.

In addition to the argument above about the nature of the algebra $\mathcal{A}_{R,0}$, it has been also studied recently by Leutheusser and Liu in \cite{leutheusser2021causal, leutheusser2021emergent} purely in the boundary theory without requiring the duality with the bulk theory.
Using the half-sided modular inclusion, they argued that above the Hawking-Page temperature, there is an emergent operator algebra which is a von Neumann algebra of Type \RNum{3}$_1$.
Take $B$ to be a time band in the right boundary and denote the algebra generated
by subtracted single-trace operators in $ B$ as $ \mathcal{A}^B_{R,0}$
which
is dual to the bulk algebra of operators in the causal wedge of the time band $B$. 
Since the generator of the boundary time translation is not part of the algebra $ \mathcal{A}_{R,0}$, the algebra $ \mathcal{A}^B_{R,0}$ is not coincide with the algebra of subtracted single-trace operators on the entire boundary and rather $\mathcal{A}^B_{R,0}$ is just a subalgebra of $\mathcal{A}_{R,0}$. 
As $ \hat{H}$ is the generator of the boundary time translation, the modular flow of the algebra $ \mathcal{A}_{R,0}$ shifts the boundary time $ t \rightarrow t+\beta u$. Therefore, the operator algebra in the time band $ B =(t_0, \infty)$ maps to itself under conjugation by $ \Delta ^{iu}$ for $ u >0$
\begin{equation}
    \Delta ^{iu} ~ \mathcal{A}^B_{R,0}~ \Delta ^{-iu}= \mathcal{A}^B_{R,0} \qquad \quad u>0. 
\end{equation}
This structure is called half-sided modular inclusion and exists only if $ \mathcal{A}_{R,0}$ is a Type \RNum{3}$_1$ von Neumann algebra.

The $1/N$ corrections to this picture have been discussed by Witten in \cite{witten2022gravity}. In particular, he showed that it modifies the emergent Type \RNum{3}$_1$ algebra to an algebra of Type \RNum{2}$_\infty$.
At large $N$ limit, one can define algebra $ \mathcal{A}_R$ as an extension of the algebra $ \mathcal{A}_{R,0}$ by adding an additional generator $U$ as 
\begin{equation}
    \mathcal{A}_R = \mathcal{A}_{R,0} \otimes \mathcal{A}_U.
\end{equation}
The algebra $ \mathcal{A}_R$ is no longer a factor since $U$ is central. Similarly, one can define the algebra $ \mathcal{A}_L$ on the left CFT by defining the operator 
$ U'= H_L'/N$.   
Note that the operators $U$ and $U'$ are not completely independent since $ U-U' = \hat{H} /N$. At strict large $N$ limit, $ U-U'$ vanishes, and therefore, the algebra $ \mathcal{A}_L$ can also be defined in terms of $U$ as 
\begin{equation}
    \mathcal{A}_L = \mathcal{A}_{L,0} \otimes \mathcal{A}_U.
\end{equation}
In the large $N$ limit, the algebras $ \mathcal{A}_L$ and $ \mathcal{A}_R$ are of Type \RNum{3}$_1$ von Neumann algebra \cite{leutheusser2021causal, leutheusser2021emergent}. Once we go beyond the large $N$ limit and consider $1/N$ corrections, the algebras $\mathcal{A}_L$ and $ \mathcal{A}_R$
become of Type \RNum{2}$_\infty$. Mathematically, the Type \RNum{2}$_\infty$ algebra is the crossed product of the Type \RNum{3}$_1$ algebra of the strict large $N$ limit by its modular automorphism group. By duality, these boundary algebras are dual to the bulk algebras of observables on each side of the black hole denoted by $\mathcal{A}_l$ and $ \mathcal{A}_r$
\begin{equation}
    \mathcal{A}_r = \mathcal{A}_R, \quad \qquad \mathcal{A}_l= \mathcal{A}_L
\end{equation}
which incorporates the algebra $ \mathcal{A}_{r(l),0}$, the observable $U$ that is central at large $N$ limit and $1/N$ corrections. Beyond $ N=\infty$, $ U$ is no longer central and the $ 1/N$ corrections modify the algebra in such a way that its center becomes trivial. More precisely perturbatively in $1/N$, the algebra of observables deforms to the factor of Type \RNum{2}$_\infty$.

\subsection{One-sided black holes in AdS/CFT}\label{3.4}

The full AdS-Schwarzschild geometry in (\ref{11}) describes an additional asymptotically AdS region that is connected to our universe by a wormhole. 
Besides them,  like in flat space, black holes can also be created by some sort of collapsing shell. In such a case, there is no wormhole connecting to another universe since the geometry at the earlier time looks nothing like the full AdS-Schwarzschild geometry as the interior is non-vacuum. Nevertheless, the one-sided geometry may share some features such as singularity and future horizon with maximally extended solutions. 

Here, there is a new important feature in comparison with flat space. The Hawking radiation in AdS black hole background will reach the boundary in a finite time and then, it is reflected back by the boundary.
If the black hole is small enough, the entire black hole evaporates before the radiation gets to the boundary. 
By increasing the size of the black hole, while the Schwarzschild radius of the black hole reaches the AdS radius, the radiation will be reflected back into the black hole very fast.
So, the black hole will quickly reach equilibrium with the Hawking radiation and remain constant in size up to small fluctuation.
As a result, with the usual reflecting boundary condition, big black holes in AdS never evaporate and they are eternal.
Thus, for a big black hole in AdS formed from collapse, at a late enough time when all the matter has fallen into the black hole and the fluctuations of the horizon have decayed away, the quantum fields start behaving like ones in the eternal black hole background.
It is known that the small black holes in AdS are not stable while the big ones are.

\begin{figure}[h]
    \centering
    
 \begin{tikzpicture}   

\draw[thick,] (3,-3)--(3,3);
\draw[thick] (-3,-3)--(-3,3);
\draw[darkgray, dashed, very thick](-3,-3) -- (3,3);
\draw[darkgray, dashed, very thick](3,-3) -- (-3,3);

\draw[red, snake=snake, segment amplitude=.4mm, segment length=2mm, thick] (-3,3)--(3,3);
\draw[red, snake=snake, segment amplitude=.4mm, segment length=2mm, thick]  (-3,-3)--(3,-3);
\node at (2,0) {${\RNum{1}}$};
\node at (0,2) {${\RNum{2}}$};
\node at (-2,0) {${\RNum{3}}$};
\node at (0,-2) {${\RNum{4}}$};

\node at (3.1,3.3) {$U=0$};
\node at (-3.1,3.3) {$V=0$};
\end{tikzpicture}\qquad\qquad
\begin{tikzpicture}
\draw[darkgray, dashed, very thick](0,-2) -- (5,3);
\filldraw[DarkSlateGray4, very thick] (5,-2) -- (0,3)--(0, 2.9) --(5, -2.1)-- cycle;
\draw[red, snake=snake, segment amplitude=.4mm, segment length=2mm, thick] (0,3)--(5,3);
\draw[thick] (0,-3) -- (0,3);
\draw[thick] (5,-3) -- (5,3);

\node at (4.5,1.5) {${\RNum{1}}$};
\node at (3,2) {${\RNum{2}}$};

\end{tikzpicture}
   
\caption{The two-sided eternal Black hole and the one-sided geometry for a stable AdS black hole created by some sort of collapsing shell.}   
\label{one-sided}
\end{figure}
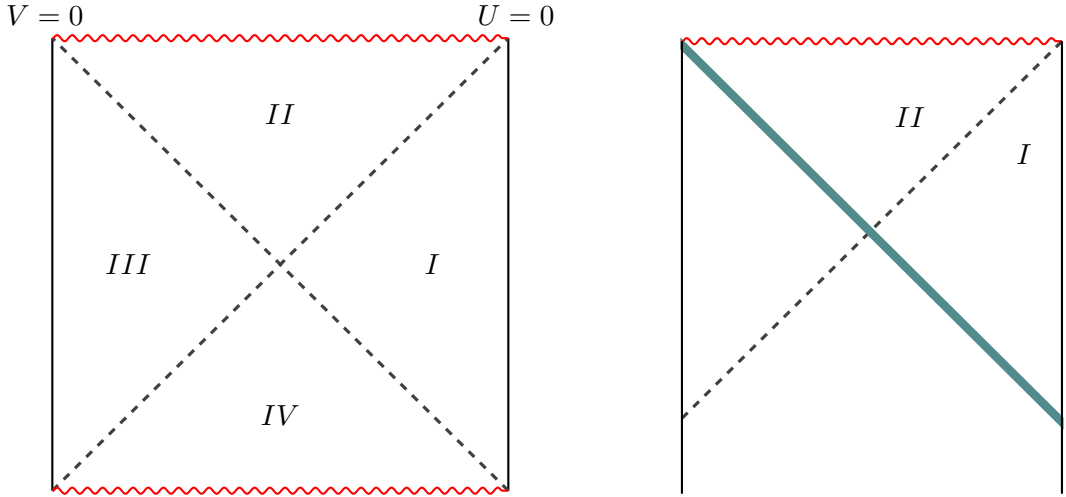

Black hole formation by collapse has a holographic interpretation as the thermalization of the CFT pure state on the boundary of the AdS space. In the dual CFT, we start in a pure state and then allow it to settle down and thermalize after a while. It will evolve to a state that is indistinguishable from a thermal state for the set of interesting observables.
Therefore, the late-time CFT correlation functions on a massive pure state can be approximated by correlation functions on the thermal density matrix.

As we said, a big black hole in a pure state is dual to a high-energy pure state in the CFT. For a black hole at fixed energy, as Bekenstein proposed, the number of black hole microstates is counted by black hole entropy $S$. At sufficiently large energy, the black hole microstates dominate the microcanonical ensemble of the CFT. In other words, almost all high-energy states in the CFT have a bulk description as a single black hole. 
In general, we can think of an equilibrium pure state as a typical state.
A typical black hole microstate of energy $E_0$ in the CFT is defined as a pure state which is a random superposition of energy eigenstates in a narrow energy band 
\begin{equation}\label{55}
    \ket{\Psi_0} = \sum _{E_i \in (E_0 - \delta E , E_0 + \delta E)} c_i \ket{E_i},
\end{equation}
where $c_i$ are random numbers selected with the uniform Haar measure.
These typical states represent the majority of black hole microstates of a given energy. They are approximately in equilibrium and so, 
it is expected that correlators in these states will be the same as thermal correlators
\begin{equation}\label{12}
    \bra{\Psi_0} O(x_1)O(x_2)...O(x_n) \ket{\Psi_0} = \frac{1}{Z_\beta} \tr( e^{-\beta H} O(x_1)O(x_2)...O(x_n))
\end{equation}
where $\beta^{-1}$ is the temperature corresponding to the energy $E_0$. We note that these typical states are not exactly the same as the late-time configuration of a black hole forming by collapse, as they have a narrower energy band.

\begin{figure}[h]
\centering
 
\begin{tikzpicture}  
\draw[thick] (3,-3)--(3,3);
\draw [darkgray, dashed,thick](-3,-3) ..controls (-2,0)..(-3,3);
\draw[darkgray](-3,-3) -- (3,3);
\draw[darkgray](3,-3) -- (-3,3);
\draw[red, snake=snake, segment amplitude=.4mm, segment length=2mm, thick] (-3,3)--(3,3);
\draw[red, snake=snake, segment amplitude=.4mm, segment length=2mm, thick]  (-3,-3)--(3,-3);
\end{tikzpicture}\qquad \qquad \qquad
\begin{tikzpicture}[>=stealth]
\draw (3,-3)--(3,3);
\draw [->](3.2,-1.5)--(3.2,1.5);
\draw[red,thick] (0,0)--(3,3);
\draw[red,thick] (0,0)--(3,-3);
\node at (3.4,0) {$t$};
\end{tikzpicture}

\caption{Two proposals for the geometry dual to the typical black hole microstate.}
\label{fire}
\end{figure}
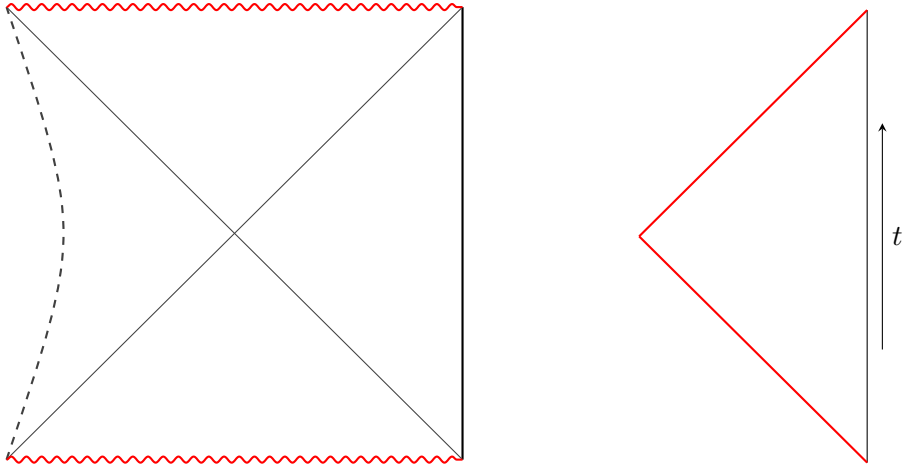
Since these typical states look time-independent for the simple observables, their dual geometry should be characterized by an approximate killing isometry which is timelike near the horizon. It is mostly accepted that the geometry contains one exterior region described by the AdS-Schwarzschild metric. 
It was proposed in \cite{almheiri2013black, almheiri2013apologia, marolf2013gauge} that the entire dual geometry is just the exterior region terminates on the horizon by a firewall which is consistent with the time translation symmetry we have.
However, since the curvature near the horizon of a big black hole is low, their proposal demands a modification of general relativity at low curvature. 
In addition to this solution,  it has been conjectured in \cite{deBoer:2018ibj, DeBoer:2019yoe} that if we have a smooth horizon, the dual geometry to a typical pure state contains the black and white hole interiors and part of the left region as well (Fig. \ref{fire}).

Finally, we mentioned that to study the evaporation of stable black holes in AdS, one can impose the absorbing boundary condition for the big black holes instead of reflecting ones. Here instead, the Hawking radiation never returns to the black hole since the outgoing modes are absorbed by the boundary, and so, the black hole evaporates.
In the dual theory then, the CFT is not a closed system and it does not evolve unitarily. However, one can as usual add an auxiliary system which here stores the outgoing Hawking radiation when it reaches the boundary.

\section{Black hole exterior reconstruction}\label{petz,bh}

In this section, we will discuss the reconstruction of the modes on the two-sided black hole background. First, we review the HKLL procedure and after that, we will explain how we can use the Petz map definition in modular theory to reconstruct the modes on the left exterior from the reconstruction of the modes on the right exterior of the black hole. 

\subsection{Reconstruction of the black hole exterior using HKLL map}

At large N we can treat the bulk theory as a quantum field theory on a curved spacetime background.  One can then represent the black hole exterior in terms of the CFT operators using the HKLL reconstruction procedure \cite{hamilton2006local, hamilton2007local, hamilton2006holographic, hamilton2008local}.
It is known that the free scalar field $\phi$ in the bulk is dual to the scalar conformal primary of the boundary with conformal dimension $ \Delta= d/2 + \sqrt{m^2 + d^2/4}$, which is related to the boundary limit of the field $\phi$ via extrapolate dictionary as
\begin{equation} \label{40}
    \lim _{r \rightarrow \infty} r^\Delta \phi (t, r, \Omega) = O (t,\Omega).
\end{equation}
In case we have a gauge theory in the boundary, these primary operators are usually some single trace operators.

Consider scalar conformal primary operator $O$.
The same as for the vacuum, large N factorization holds for the thermal correlation functions, i.e.
\begin{multline}
        \tr\big(\rho_{th} O(x_1)...O(x_{2n})\big) = 
        \\
        \frac{1}{2^n} \sum _\pi 
      \tr \big( \rho_{th} O(x_{\pi_1})O(x_{\pi_2})\big) ...   \tr\big( \rho_{th} O(x_{\pi_{2n-1}})O(x_{\pi_{2n}})\big) + O(1/N),
\end{multline}
where $\pi$ runs over the set of permutations.
From (\ref{12}), one can find out that the large N factorization holds for the typical pure states as well, thus in all cases, each Schwarzschild mode in the bulk is dual to a  GFF on the boundary.

We can expand the boundary GFF in terms of its Fourier modes $ O_{\omega,m}$ as
\begin{equation}\label{13}
    O(t,\Omega) = \sum _m \int_0^{\infty} \frac{d\omega}{2\pi}
    \big(g_{\omega,m}(t, \Omega)O_{\omega,m} + g^*_{\omega,m}(t,\Omega)O^\dagger_{\omega,m}\big).
\end{equation}
The thermal expectation values of the Fourier operators also imply that they behave like the unnormalized creation and annihilation operators. One can use the extrapolate dictionary to find the rescaled operators $ \hat{O}_{\omega,m}= M_{\omega,m}^{-1} O_{\omega,m}$, which are identified with the bulk modes $ a_{\omega,m}$.
These CFT operators $ \hat{O}_{\omega,m}$ are the ones thermally populated at the Hawking temperature of the black hole $ \beta^{-1}$
\begin{equation}
  \begin{split}
        \frac{1}{Z_\beta} \tr (e^{-\beta H} \hat{O}_{\omega,m}\hat{O}_{\omega',m'}^\dagger)
        = &\frac{e^{\beta\omega}}{e^{\beta\omega}-1} \delta(\omega-\omega')\delta_{m,m'}
        \\
        \frac{1}{Z_\beta} \tr(e^{-\beta H} \hat{O}^\dagger_{\omega,m}\hat{O}_{\omega',m'}) = &\frac{1}{e^{\beta\omega}-1} \delta(\omega-\omega')\delta_{m,m'}.
    \end{split}  
\end{equation}
Having the identification between bulk and boundary modes, we can follow the mode sum approach in \cite{hamilton2007local} and find the CFT expression for every bulk field outside the horizon as 
\begin{equation}\label{14}
    \Phi _{HKLL} (t,r,\Omega) = \int dt' d\Omega'~ K(t,r,\Omega| t',\Omega') O(t',\Omega')
\end{equation}
for an appropriate choice of smearing function $K$. We can also find the field expression in terms of Fourier modes by plugging (\ref{13}) into (\ref{14}) as
\begin{equation}\label{17}
    \Phi _{HKLL} (t,r,\Omega) =  \sum _m \int_0^{\infty} \frac{d\omega}{2\pi}
     \big(\mathcal{F}_{\omega,m}(t,r, \Omega)O_{\omega,m} + \mathcal{F}^*_{\omega,m}(t,r,\Omega)O^\dagger_{\omega,m}\big)
\end{equation}
where 
\begin{equation}
    \mathcal{F}_{\omega,m}(t,r, \Omega)=
    \int dt' d\Omega'~ K(t,r,\Omega| t',\Omega') g_{\omega,m}(t',\Omega'). 
\end{equation}
By comparing with (\ref{8}), we can find that $ \mathcal{F}_{\omega,m}(t,r, \Omega)= M^{-1}_{\omega,m} f_{\omega,m}(t,r,\Omega) $.

\subsection{Coarse-grained vs fine-grained observables}\label{4.2}

Consider a big AdS black hole in equilibrium. An observable outside the horizon of the black hole has access just to the information in the exterior of the black hole, referred to as region \RNum{1}. This bulk observable can not distinguish the microstate of the black hole and more generally, the fact that it is a one-sided black hole or connected to another universe through a wormhole.  
However, in all cases, if we are just interested to the low energy observables outside the horizon, we find from \eqref{15}
that it is enough to describe the system by a thermal density matrix.

On the other hand, having a big black hole in AdS is dual to the thermalization of the boundary theory. In general, the thermalization of a closed quantum system leads to the division of the observables of the theory into two parts:
the coarse-grained or macroscopic observables and the fine-grained or microscopic ones, denoted by $ \mathcal{A}_{c}$ and $ \mathcal{A}_{f}$, respectively.
The coarse-grained observables are the ones that can be easily measured by the low-energy observer.
More precisely, the thermalization of the system means that if we are just interested in measuring the macroscopic observable, the system can be approximately described by a thermal density matrix, $i.e.$  
\begin{equation}\label{19}
    \rho\big|_{\mathcal{A}_{c}} = \rho_{th, c} = \frac{1}{Z^{c}_\beta} e^{-\beta H_{c}},
\end{equation}
where $H_c$ is the coarse-grained Hamiltonian and $Z^{c}_\beta$ is the coarse-grained partition function
\cite{papadodimas2013infalling}.

In AdS/CFT where we have the duality between the bulk and boundary theories, the bulk Hilbert space is isomorphic to the boundary one. In the bulk, we have a fundamental theory of quantum gravity that in low energy described
by a local quantum field theory on a curved spacetime background. These are usually the macroscopic degrees of freedom in the bulk while the stringy and trans-Planckian observables are the non-perturbative microscopic degrees of freedom.  
When we have a black hole in the bulk, the coarse-grained observables are just the operators that lie outside the horizon while the fine-grained one contains the degrees of freedom of the black hole interior as well as the non-perturbative ones on the entire bulk.
In the rest, we are interested in studying the bulk gravity in the low energy and we denote the algebra of operator in this regime in the exterior and interior of the black hole by $ \mathcal{A}_{ext}$ and $ \mathcal{A}_{int}$, respectively.

As it is mentioned above, for the low-energy observables outside the horizon
we can describe the bulk theory by the thermal density matrix for the bulk effective field theory lives in the AdS-Schwarzschild coordinates, in other words
\begin{equation}\label{20}
    \rho_{bulk}\Big|_{\mathcal{A}_{ext}} = \rho_{th, \RNum{1}}. 
\end{equation}

\subsection{Reconstruction of the black hole exterior using the Petz map}\label{4.3}

As previously described, one can map the local bulk field in the exterior of a black hole  into the non-local CFT operators using the HKLL procedure
\begin{equation}
    \phi (t,r,\Omega) \longrightarrow \Phi _{HKLL} (t,r,\Omega) = \int_{bdy} dt' d\Omega' ~K(t,r,\Omega| t',\Omega') O(t',\Omega').
\end{equation}
In other words, the HKLL map provides us an isometry of embedding $ V_{HKLL}: \mathcal{H}_{ext} \rightarrow \mathcal{H}_{CFT}$ which in the Heisenberg picture  maps the operators as
\begin{equation}
    \phi (t,r,\Omega) \longrightarrow \Phi _{HKLL} (t,r,\Omega)= V_{HKLL}~\phi (t,r,\Omega) ~V_{HKLL}^\dagger .
\end{equation}
It is equivalent to consider the quantum channel which maps the density matrices in the exterior region to the boundary density matrices $ \mathcal{E}: \mathcal{S}(\mathcal{A}_{ext}) \rightarrow \mathcal{S}(\mathcal{A}_{CFT})$ as
\begin{equation}
   \mathcal{E}(.) = V_{HKLL}~(.)~ V^\dagger_{HKLL}.
\end{equation}
However, we should be careful that the low-energy observers can only measure the coarse-grained operators, for any GFF $O(t, \Omega)$, they can measure just
\begin{equation}
   O_c(t, \Omega) = P_{coarse} O(t, \Omega) P_{coarse}
\end{equation}
where $ P_{coarse}$ is the projection onto the coarse-grained Hilbert space which traces out the fine-grained degrees of freedom. Thus, the actual map we have is
\begin{multline}
    \phi (t,r,\Omega) \longrightarrow P_{coarse}~\Phi _{HKLL} (t,r,\Omega) ~ P_{coarse}
    \\
   =P_{coarse}~ V_{HKLL}~\phi (t,r,\Omega)~ V_{HKLL}^\dagger ~P_{coarse}
   =\int_{bdy} dt' d\Omega'~ K(t,r,\Omega| t',\Omega') O_c(t',\Omega').
\end{multline}
The Hilbert space of these coarse-grained GFF has the Fock space structure which is isomorphic to the Hilbert space of the free fields on the AdS-Schwarzschild background. Therefore, we can introduce the quantum channel 
$ \mathcal{E}_c: \mathcal{S}(\mathcal{A}_{ext}) \rightarrow \mathcal{S}(\mathcal{A}_{coarse}) $ as
\begin{equation}
     \mathcal{E}_c(.) = V_{HKLL}~(.)~ V^\dagger_{HKLL} \Big|_{\mathcal{A}_{coarse}} = V_{c}~(.)~ V^\dagger_{c}
\end{equation}
where $ V_c = P_{coarse} V_{HKLL} $. Unlike $ \mathcal{E}$, the quantum channel $ \mathcal{E}_c$ is invertible as the evolution is done via a unitary evolution. In this case, the recovery channel is simply known as 
\begin{equation}\label{56}
    \mathcal{R}_c(.) = V^\dagger_{c}~(.) ~V_{c} =  \mathcal{E}^*_c(.),
\end{equation}
which one can also find using the Petz recovery channel formula \eqref{14}.
Therefore, one can use the dual of the recovery channel to map the operators in the Heisenberg picture, $ \mathcal{R}_c^* :\mathcal{A}_{ext} \rightarrow \mathcal{A}_{coarse} $ 
\begin{equation}
     \mathcal{R}^*_c(.) = V_{c}~(.)~ V_{c}^\dagger=  \mathcal{E}_c(.).
\end{equation}
\begin{figure}[h]
    \centering
    
    \begin{tikzpicture}[>=stealth]
    
    \node at ( 0,0) (coarse) {$\mathcal{S}(\mathcal{A}_{coarse})$};
    \node at ( 0,3) (ext) {$\mathcal{S}(\mathcal{A}_{ext})$};
    \node at (4,0)  (CFT) {$\mathcal{S}(\mathcal{A}_{CFT})$};
    \draw [->] (CFT.west) -- (coarse.east);
    \draw [->, dashed] (0,2.6) -- (0,0.3);
    \draw [->] (0.5,2.6) -- (3.8,0.4);
    \node at ( 2 ,-0.4)  {$res$};
    \node at (2.8, 1.5 )  {$\mathcal{E}$};
    \node at (-0.5, 1.5 )  {$\mathcal{E}_c$};
    
    \node at ( 8,0)  {$\mathcal{A}_{coarse}$};
    \node at ( 8,3)  {$\mathcal{A}_{ext}$};
    \draw [->, dashed] (8,2.6) -- (8,0.3);
    \node at ( 8.5 ,1.5)  {$\mathcal{R}_{c}^*$};
    \end{tikzpicture}

   
\end{figure}
As a result, we have
\begin{equation}
     \mathcal{R}^*_c\big( \phi (t,r,\Omega)\big)=\int_{bdy} dt' d\Omega'~ K(t,r,\Omega| t',\Omega') O_c(t',\Omega')
\end{equation}
From now on, we drop the subscript $c$, but we mean by $O$ the coarse-grained GFF.
Since 
$ \mathcal{A}_{ext} = \text{span} \{a_{\omega,m}, a^\dagger_{\omega,m}\}$, it is enough to find the action of the recovery map $ \mathcal{R}_{c}^*$ on these set of operators.
One can easily use \eqref{8}, \eqref{17} and reach to
\begin{equation}\label{57}
    \mathcal{R}^*_c(a_{\omega,m}) = \hat{O}_{\omega,m}, \qquad  \mathcal{R}^*_c(a^\dagger_{\omega,m}) = \hat{O}^\dagger_{\omega,m}.
\end{equation}

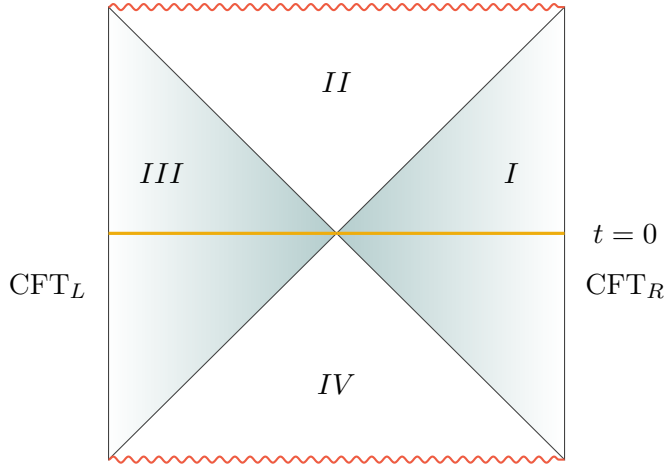
\begin{figure}[h]
    \centering
    
    \begin{tikzpicture}
        \shadedraw[left color= LightCyan3,right color=white, draw=darkgray](0,0)--(3,3)--(3,-3)--cycle;
        \shadedraw[right color= LightCyan3,left color=white, draw=darkgray](0,0)--(-3,3)--(-3,-3)--cycle;     
        \draw[Tomato2, thick, snake=snake, segment amplitude=.4mm, segment length=2mm] (-3,3)--(3,3);
        \draw[Tomato2, thick, snake=snake, segment amplitude=.4mm, segment length=2mm]  (-3,-3)--(3,-3);

        \node at (2.3,0.8) {${\RNum{1}}$};
        \node at (0,2) {${\RNum{2}}$};
        \node at (-2.3,0.8) {${\RNum{3}}$};
        \node at (0,-2) {${\RNum{4}}$};

        \draw[DarkGoldenrod2, very thick] (-3,0) -- (0,0);
        \draw[DarkGoldenrod2, very thick] (0,0) -- (3,0);
        \node at (3.8,-0.7) {CFT$_{R}$};
        \node at (-3.8,-0.7) {CFT$_{L}$};
        \node at (3.8,0) {$t=0$};

    \end{tikzpicture}

\caption{The two-sided eternal black hole in holography. }
\label{eternal2}
   
\end{figure}
For now, let us consider a two-sided geometry that is dual to the TFD state of the two identical non-interacting CFTs in the boundary which is given by \eqref{21},
and take the $t=0$ Cauchy slice in the bulk, Fig. \ref{eternal2}. 
As we had in Sec. \ref{sec2}, the bulk Hilbert space corresponding to the quantizing the small fluctuations around the black hole geometry is denoted as $ \mathcal{H}^{(Fock)}_{BH}$ and the algebra of low-energy observables on two sides of the black hole as $ \mathcal{A}_{l,0}$ and $ \mathcal{A}_{r,0}$. The bulk Hilbert space can be constructed through the action of the operator algebra of only the right exterior on the HH state (one can obtain it equivalently from the algebra of the left exterior), i.e.
\begin{equation}
    \mathcal{H}^{(Fock)}_{BH} = \mathcal{A}_{r,0} \ket{HH} = \mathcal{A}_{l,0} \ket{HH}.
\end{equation}
The Hilbert space of the full boundary theory  is
\begin{equation}
    \mathcal{H} = \mathcal{H}_{CFT_L} \otimes \mathcal{H}_{CFT_R}.
\end{equation}
The bulk states in the Hilbert space $\mathcal{H}^{(Fock)}_{BH}$ are dual to the set of states called \emph{code subspace} in the boundary theory corresponding to the excitations around the TFD state. They can be obtained by acting with the dual single-trace boundary operators on the TFD state denoted by $ \mathcal{H}_{TFD}$.
The TFD code subspace spanned by the states $ a \ket{\Psi_{TFD}}$ with $ a\in \mathcal{A}_{R,0}$.
The corresponding code subspace has a structure of a Hilbert space that can be made via GNS construction by using only $\mathcal{A}_{R,0}$ or $ \mathcal{A}_{L,0}$ over TFD state (We will discuss it more precisely later).
At large $N$ limit where the algebras are of Type \RNum{3}$_1$, the TFD Hilbert space does not have a tensor product structure, in other words, there are no candidates for $ \mathcal{H}_L$ and $ \mathcal{H}_R$ such that 
$ \mathcal{H}_{TFD} = \mathcal{H}_L \otimes \mathcal{H}_R$. 
The GFFs are then the representations of the single trace operators on the TFD Hilbert space and thus the algebras $ \mathcal{A}_{L,0}$ and $ \mathcal{A}_{R,0}$ are the von Neumann algebras on $ \mathcal{H}_{TFD}$. 
We note that these representations are not exactly the same as the original operators since they only define on $ \mathcal{H}_{TFD}$ while the single trace operators act on the full CFT Hilbert space. 

We can follow the discussion above and map the algebra of operators in each region to the coarse-grained operators on the boundary via two copies of the Petz recovery channel as
\begin{equation}
        \mathcal{R}^*_{c,R}: \mathcal{A}_{r,0} \longrightarrow \mathcal{A}_{R,0},  \qquad\qquad\mathcal{R}^*_{c,L}: \mathcal{A}_{l,0 }\longrightarrow \mathcal{A}_{L,0}
\end{equation}
such that 
\begin{equation}
    \begin{split}
        \mathcal{R}&^*_{c,R}(a_{\omega,m}) = \hat{O}_{\omega,m;R}, \qquad  \mathcal{R}^*_{c,R}(a^\dagger_{\omega,m}) = \hat{O}^\dagger_{\omega,m;R}
        \\
        \mathcal{R}&^*_{c,L}(\Tilde{a}_{\omega,m}) = \hat{O}_{\omega,m;L}, \qquad  \mathcal{R}^*_{c,L}(\Tilde{a}^\dagger_{\omega,m}) = \hat{O}^\dagger_{\omega,m;L}.
    \end{split}
\end{equation}
Knowing the boundary reconstruction of the Schwarzschild modes, we can find the boundary representation of the bulk field in every bulk point.
Although we can simply map the operator algebra in region \RNum{3} to the left CFT through $ \mathcal{R}_{c,L}^*$, there is an alternative way to write this mapping by using the Petz map formula in the GNS Hilbert space \eqref{7}. This will be helpful later when we are interested in finding the CFT representation of the one-sided black hole interior modes.

As we mentioned, the effective field theory on the eternal black hole background is described by the HH state, thus
\begin{equation}
  \rho_{bulk}\Big|_{\mathcal{A}_{r,0}} = \rho_{th, \RNum{1}} \qquad\qquad\qquad \rho_{bulk}\Big|_{\mathcal{A}_{l,0}} = \rho_{th, \RNum{3}}, 
\end{equation}
and it is cyclic and separating for the operator algebra both in regions \RNum{1} and \RNum{3}. Moreover,
since these two regions are spacelike separated, we have 
$ [\mathcal{A}_{l,0}, \mathcal{A}_{r,0}]=0$ and so they are each others commutants
\begin{equation}
   ( \mathcal{A}_{l,0})' = \mathcal{A}_{r,0}.
\end{equation}
On the other hand, the boundary theory is in the TFD state, therefore each CFT itself is described by the CFT thermal state and as we mentioned above, its restriction to the coarse-grained algebra is also a thermal state but this time with respect to the coarse-grained Hamiltonian. 
They are also the commutants of each other on the $ \mathcal{H}_{TFD}$
\begin{equation}
   (\mathcal{A}_{L,0})'=\mathcal{A}_{R,0}.
\end{equation}
Therefore, one can use the modular theory expression \eqref{7} and write the Petz map $ \mathcal{R}_{c,L}^*$ as 
\begin{equation}
    \mathcal{R}_{c,L}^*(.) = \mathcal{J}_{TFD} \circ \mathcal{R}'_{c,R-\rho_{th}} \circ \mathcal{J}_{HH} ~(.) = J_{TFD}~ \mathcal{R}'_{c,R-\rho_{th}}~ (J_{HH}~ (.)~J_{HH} )~ J_{TFD}
\end{equation}
while $ \mathcal{J}_{HH}$ and $ \mathcal{J}_{CFT}$ are respectively the modular conjugations of the bulk and boundary theories with respect to the low-energy observables, and, $\mathcal{R}'_{c,R-\rho_{th}} $ is defined based on the relation in \eqref{6}.
\begin{figure}[h]
    \centering
 
    \begin{tikzpicture}[>=stealth]
    
    \node at ( 0,0)  {$\mathcal{A}_{L,0}$};
    \node at ( 0,3)  {$\mathcal{A}_{l,0}$};
    \draw [->, dashed] (0,2.6) -- (0,0.3);
    \node at ( -0.5 ,1.5)  {$\mathcal{R}_{c,L}^*$};
    
    \node at ( 5,0) (L) {$\mathcal{A}_{L,0}$};
    \node at ( 5,3) (3) {$\mathcal{A}_{l,0}$};
    
    \node at ( 9,0)  (R){$\mathcal{A}_{R,0}$};
    \node at ( 9,3)  (1){$\mathcal{A}_{r,0}$};
    \draw [->, dashed] (9,2.6) -- (9,0.3);
    \node at ( 9.9 ,1.5)  {$\mathcal{R}'_{c,R-\rho_{th}}$};
    
    \draw [->] (R.west) -- (L.east);
    \draw [->] (3.east) -- (1.west);
    
    \node at ( 7,3.5)  {$\mathcal{J}_{HH}$};
    \node at ( 7,-0.5)  {$\mathcal{J}_{TFD}$};
  
    \end{tikzpicture}

   
\end{figure}

In the bulk where the theory is described by  HH state, the modular conjugation operator is the anti-unitary  CPT operator-more precisely, CRT transformation- which in the AdS-Schwarzschild coordinates acts as
\begin{equation}
\mathcal{J}_{HH} ~\big(\phi_{\RNum{1}}(t,r,\Omega)\big)= J_{HH}~ \phi_{\RNum{1}}(t,r,\Omega) ~J_{HH} = \phi_{\RNum{3}}(-t,r,\Omega).
\end{equation}
One then can find that 
\begin{equation}
    \mathcal{J}_{HH} : ~ a_{\omega,m} \longleftrightarrow \Tilde{a}_{\omega,m}.
\end{equation}
On the boundary side where the theory is described by the TFD state, the modular conjugation acts in the TFD Hilbert space as
\begin{equation}
    \mathcal{J}_{TFD} : ~ O_{\omega,m;L} \longleftrightarrow  O_{\omega,m;R}.
\end{equation}
Therefore, one can simply check that 
\begin{equation}
    \begin{split}
        \mathcal{R}&_{c,L}^*(\Tilde{a}_{\omega,m}) = \mathcal{J}_{TFD} \circ \mathcal{R}'_{c,R-\rho_{th}} \circ \mathcal{J}_{HH} (\Tilde{a}_{\omega,m}) = \hat{O}_{\omega,m;L}
        \\
        \mathcal{R}&_{c,L}^*(\Tilde{a}_{\omega,m}^\dagger) = \mathcal{J}_{TFD} \circ \mathcal{R}'_{c,R-\rho_{th}} \circ \mathcal{J}_{HH} (\Tilde{a}_{\omega,m}^\dagger) = \hat{O}^\dagger_{\omega,m;L}.
    \end{split}
\end{equation}

Before going ahead we note here that after considering the $ 1/N$ corrections, the picture needs some modifications. 
The algebra of observables on each side of the eternal black hole is $ \mathcal{A}_l$ and $ \mathcal{A}_r$ which are of Type \RNum{2}$_\infty$. They are dual to the crossed product of  $ \mathcal{A}_{L,0}$ and $ \mathcal{A}_{R,0}$
and their group of modular automorphism. The mapping now is as below.
\begin{figure}[h]
    \centering
 
    \begin{tikzpicture}[>=stealth]

    \node at ( -2,0)  {$\mathcal{A}_{L}$};
    \node at ( -2,3)  {$\mathcal{A}_{l}$};
    \draw [->, dashed] (-2,2.6) -- (-2,0.3);
    \node at ( -2.5 ,1.5)  {$\mathcal{R}_{c,L}^*$};
 
    \node at ( 1,0)  {$\mathcal{A}_{r}$};
    \node at ( 1,3)  {$\mathcal{A}_{R}$};
    \draw [->, dashed] (1,2.6) -- (1,0.3);
    \node at ( 1.5 ,1.5)  {$\mathcal{R}_{c,R}^*$};
    
    \node at ( 5,0) (L) {$\mathcal{A}_{L}$};
    \node at ( 5,3) (3) {$\mathcal{A}_{l}$};
    
    \node at ( 9,0)  (R){$\mathcal{A}_{R}$};
    \node at ( 9,3)  (1){$\mathcal{A}_{r}$};
    \draw [->, dashed] (9,2.6) -- (9,0.3);
    \node at ( 9.9 ,1.5)  {$\mathcal{R}'_{c,R-\rho_{th}}$};
    
    \draw [->] (R.west) -- (L.east);
    \draw [->] (3.east) -- (1.west);
    
    \node at ( 7,3.5)  {$\mathcal{J}_{HH}$};
    \node at ( 7,-0.5)  {$\mathcal{J}_{TFD}$};
  
    \end{tikzpicture}

   
\end{figure}
It should be considered that since we are working at  $1/N$ correction, it is better even to use the approximate version of the recovery channel introduced in Sec. \ref{approx}.

\subsection{Vacuum of the GNS Hilbert space}\label{4.4}

As we had in the previous section, the Hilbert space of the effective field theory in bulk is dual to the TFD Hilbert space $ \mathcal {H}_{TFD}$.
In this section, we will discuss more precisely the structure of dual code subspace on the boundary.

The boundary theory is the tensor factor of two identical CFTs, $ \mathcal{H} = \mathcal{H}_{CFT_L} \otimes \mathcal{H}_{CFT_R}$. The algebra of bounded operators on $ \mathcal{H}$ is 
\begin{equation}
   \mathcal{L}(\mathcal{H}) = \mathcal{L}(\mathcal{H}_{CFT_L}) \otimes \mathcal{L}(\mathcal{H}_{CFT_R}) 
\end{equation}
and the algebra of low-energy observables is a subalgebra of the full algebra
\begin{equation}
    \mathcal{A}= \mathcal{A}_{L,0} \otimes \mathcal{A}_{R,0}.
\end{equation}
In order to define the TFD Hilbert space as explained in \cite{leutheusser2021emergent} we should  associate a state $ \ket{a}$ to each operator $ a \in \mathcal{A}$ with the inner product among them which is defined as 
\begin{equation}
    \langle a \ket{b} = \bra{\Psi_{TFD}} a^\dagger b \ket{\Psi_{TFD}}
\end{equation}
for all $ a,b \in \mathcal{A}$ and in particular if both $ a,b$ belong to $ \mathcal{A}_{R,0}$ or $ \mathcal{A}_{L,0}$, it is reduced to 
\begin{equation}
     \langle a \ket{b} = \tr (\rho_{th} a^\dagger b).
\end{equation}
The set of vectors $ \ket{a}$ does not have a Hilbert space structure since there exists non-zero operators $ y \neq 0$ in the algebra $ \mathcal{A}$ such that $\langle y \ket{y} =0$ \cite{leutheusser2021emergent}.
In other words, the TFD state is not separating for the algebra $ \mathcal{A}$. It is just cyclic and separating for the full operator algebra on each CFT. 
In such a case to construct a Hilbert space from this set of vectors, we can use the GNS construction to set such a vectors to zero by  introducing the set of equivalence classes. The equivalence relation between them is defined as
\begin{equation}
    a \sim a+y \qquad\qquad a \in \mathcal{A},~ y \in \mathcal{Y}
\end{equation}
while $ \mathcal{Y}$ is the set of operators such that $ \langle y \ket{y} =0$. 
Moreover, since the action of the single trace operators on both sides of TFD state are related to each other,
only the algebra $ \mathcal{A}_{R,0}$ or $ \mathcal{A}_{L,0}$ is enough to generate the full TFD Hilbert space. In other words, all the vectors in $ \mathcal{H}_{TFD}$ can be written as $ \ket{a}$ with $ a \in \mathcal{A}_{R,0} $. 

Now let us consider just the algebra of single trace operators $ \mathcal{A}_{R,0}$ on the right boundary theory and build the GNS Hilbert space of the algebra with respect to the thermal density matrix which is denoted as $ \mathcal{H}_{\rho_{th}}^{ GNS}$.  
In the GNS Hilbert space, the thermal state is represented by a pure state denoted by $ \ket{ \Omega_0}$ which is called the GNS vacuum. It is also the state in $ \mathcal{H}^{GNS}_{\rho_{th}}$ corresponds to the identity operator of the algebra $ \mathcal{A}_{R,0}$.
The GNS construction provides a representation for the algebra $ \mathcal{A}_{R,0}$ on $ \mathcal{H}^{GNS}_{\rho_{th}}$ which we denote here as $ \mathcal{M}_R$.
The representation of any operator $ a \in \mathcal{A}_{R,0}$ is $ \pi (a) \in \mathcal{M}_R$ that acts only on $ \mathcal{H}^{GNS}_{\rho_{th}}$ and thus, it is \emph{state-dependent} since it depends on the state that the GNS Hilbert space is built over that. On the other hand, the original operator acts on the full CFT Hilbert space and is \emph{state-independent}.  
The inner product among the states in the GNS Hilbert space is written as
\begin{equation}
    \langle a | b \rangle = \bra{ \Omega_0} \pi (a)^\dagger \pi (b) \ket{ \Omega_0}.
\end{equation}
Here the algebra consists of the single trace operators of the CFT. Their representations on the GNS Hilbert space are the GFFs acting only on $ \mathcal{H}^{GNS}_{\rho_{th}}$ and we also have
\begin{equation}
    \bra{ \Omega_0} \pi \big(O_R(x_1)\big)^\dagger \pi \big( O_R(x_2)\big) \ket{ \Omega_0} = \tr \big( \rho_{th} O(x_1)^\dagger  O(x_2)\big).
\end{equation}

As we had, the TFD Hilbert space can be obtained using just $ \mathcal{A}_{R,0}$ or $\mathcal{A}_{L,0}$ alone over the TFD state. While the boundary theory is in the TFD state, the CFT$_R$ is described by thermal state and so there should be some relation between the TFD Hilbert space and the GNS Hilbert space corresponding to the thermal matrix over $ \mathcal{A}_{R,0}$. Indeed, it can be shown that they are isomorphic
\begin{equation}
    \mathcal{H}_{TFD}~ \cong ~\mathcal{H}^{GNS}_{\rho_{th}}.
\end{equation}
We defined the algebra $ \mathcal{M}_R$ to be the representation of the $ \mathcal{A}_{R,0}$ on the GNS Hilbert space. Therefore, its commutant which we denote as $ \mathcal{M}_L$ can be interpreted as the representation of the $ \mathcal{A}_{L,0}$ on $ \mathcal{H}^{GNS}_{\rho_{th}}$.

The GNS vacuum $ \ket{\Omega_0}$ is cyclic and separating for both $ \mathcal{M}_R$
and $ \mathcal{M}_L$.
Therefore, there is a modular operator for the algebras which generate automorphisms of them and leaves $ \ket{\Omega_0}$ invariant. If we denote the modular operator for $ \mathcal{M}_R$
as $ \Delta_0$, then $ \Delta_0^{-1}$ is the modular operator for the algebra $ \mathcal{M}_L$.
It can be seen as the representation of the $ \Delta$, the modular operator for $ \mathcal{A}_{R,0}$, in the GNS Hilbert space and in particular, it should satisfy 
\begin{equation}
    \pi \big( \Delta^{-iu} a~ \Delta^{iu} \big ) = \Delta^{-iu}_0 \pi ( a)~ \Delta^{iu} _0
\end{equation}
for all $ a \in \mathcal{A}_{R,0}$.

Now consider bulk field $ \phi$ which is dual to the boundary operator $O$ in the AdS/CFT dictionary. 
To be more precise from the algebraic point of view, we should say that the bulk field restricted in the regions \RNum{1} and \RNum{3} are dual to the representations of the $O_R$
and $ O_L$ in the GNS Hilbert space that here they are nothing but GFFs.
The extrapolate dictionary \eqref{40} should also be written more carefully as 
\begin{equation}
\begin{split}
     \pi \big( O_R(t,\Omega) \big)& = \lim _{r \rightarrow \infty} r^\Delta ~\phi_R (t, r, \Omega) 
     \\
      \pi \big( O_L(t,\Omega) \big)& = \lim _{r \rightarrow \infty} r^\Delta ~\phi_L (t, r, \Omega) .
\end{split}
\end{equation}
Under the duality, at strict large $N$ limit we reach to the identifications:
\begin{equation}
    \mathcal{H}^{GNS}_{\rho_{th}} = \mathcal{H}^{(Fock)}_{BH}, \quad \quad 
    \ket{\Omega_0} = \ket{HH}, \quad\quad \mathcal{M}_R= \mathcal{A}_{r,0},
    \quad\quad \mathcal{M}_L= \mathcal{A}_{l,0}.
\end{equation}
Moreover, the state-dependence of the GNS representations of the boundary dual operators is indeed a reflection of the fact that when we treat the gravity at weak coupling in bulk, it's mode expansion that identifies the bulk operator algebra for us depends on the bulk semi-classical geometry.

In order to create a GNS Hilbert space to describe the code subspace in the boundary, one can alternatively start with another cyclic and separating vector $ \ket{\Omega}$ for the algebra $ \mathcal{M}_R$ as the vacuum. Therefore, by duality, we have 
\begin{equation}
     \ket{\Omega} = \ket{HH}.
\end{equation}
In general, the new vacuum can be related to $ \ket{\Omega_0}$ by a unitary as 
\begin{equation}\label{47}
     \ket{\Omega} = U \ket{\Omega_0}.
\end{equation}
In particular, the simple cases are in the form of 
\begin{equation}
    \ket{\Omega} = v_L w_R \ket{\Omega_0}
\end{equation}
while $ v_L \in \mathcal{M}_L$ and  $ w_R \in \mathcal{M}_R$.
The vacuum $ \ket{\Omega_0}$ chose to be related to building the GNS Hilbert space around the TFD state of the boundary theory. Then the GNS vacuum $ \ket{\Omega}$  interpretation depends on whether it belongs to 
$ \mathcal{H} ^{GNS} _{\Omega_0}$ or not. 
If it does so, $i.e.$ 
\begin{equation}
    \mathcal{H} ^{GNS} _{\Omega} = \mathcal{H} ^{GNS} _{\Omega_0}
\end{equation}
the new GNS vacuum
corresponds to having some small excitations around the eternal black hole background
and the unitaries $ w_R$ and $ v_L$ are related to the excitations which lie just in region \RNum{1} and \RNum{3} respectively. 
On the other hand, if the state $ \ket{\Omega}$ does not lie in the GNS Hilbert space, bulk geometry is no longer described by the eternal black hole. The new vacuum can be related to excitations due to the unitary that changes the energy of the system by an amount that scales with $ N$ and thus its backreaction changes the geometry of the spacetime. 
Another possible example could be the time-shifted TFD state, defined as
\begin{equation}
    \ket{\Psi_T} = e^{i (H_L + H_R)T/2} \ket{\Psi_{TFD}} = e^{i H_LT} \ket{\Psi_{TFD}} = e^{i H_RT} \ket{\Psi_{TFD}}
\end{equation}
or more generally evolving the TFD state with some other global charges. They correspond to large diffeomorphisms in the bulk which for some large value of $T$, it changes the bulk geometry (for more detail see \cite{Papadodimas:2015jra}).

More generally we can extend the discussion in Sec. \ref{4.3} and write the mapping from the bulk algebra of observables to the representation of the corresponding algebra on the boundary as 
\begin{equation}
        \mathcal{R}^*_{c,R}: \mathcal{A}_{r,0} \longrightarrow \mathcal{M}_{R},  \qquad\qquad\mathcal{R}^*_{c,L}: \mathcal{A}_{l,0 }\longrightarrow \mathcal{M}_{L}
\end{equation}
\begin{figure}[h]
    \centering
    \begin{tikzpicture}[>=stealth]
    
    
    \node at ( 5,0) (L) {$\mathcal{M}_{L}$};
    \node at ( 5,3) (3) {$\mathcal{A}_{l,0}$};
    
    \node at ( 9,0)  (R){$\mathcal{M}_{R}$};
    \node at ( 9,3)  (1){$\mathcal{A}_{r,0}$};
    \draw [->, dashed] (9,2.6) -- (9,0.3);
    \node at ( 9.9 ,1.5)  {$\mathcal{R}'_{c,R-\Omega}$};
    
    \draw [->] (R.west) -- (L.east);
    \draw [->] (3.east) -- (1.west);
    
    \node at ( 7,3.5)  {$\mathcal{J}_{HH}$};
    \node at ( 7,-0.5)  {$\mathcal{J}_{GNS, \Omega}$};
  
    \end{tikzpicture}
\end{figure}
while again to map the operators lies in the region \RNum{3} to the left CFT, we can use the Petz map definition in the modular theory 
\begin{equation}\label{50}
    \mathcal{R}_{c,L}^*(~.~) = \mathcal{J}_{GNS, \Omega} \circ \mathcal{R}'_{c,R-\Omega} \circ \mathcal{J}_{HH} (~.~) 
\end{equation}
where the bulk mode maps as 
\begin{equation}
    \begin{split}
        \Tilde{a}_{\omega,m} & \longrightarrow \mathcal{J}_{GNS, \Omega} \circ \mathcal{R}'_{c,R-\Omega}\circ \mathcal{J}_{HH} (\Tilde{a}_{\omega,m}) = \pi\big (\hat{O}_{\omega,m;L}\big)
        \\
    \Tilde{a}^\dagger_{\omega,m} & \longrightarrow \mathcal{J}_{GNS, \Omega} \circ \mathcal{R}'_{c,R-\Omega} \circ \mathcal{J}_{HH} (\Tilde{a}^\dagger_{\omega,m}) = \pi\big (\hat{O}^\dagger_{\omega,m;L}\big)
    \end{split}
\end{equation}
and $ \pi$ provides the representation of the single trace operators in $ \mathcal{H}_\Omega^{GNS}$.

Let us consider two vacua, $ \ket{\Omega}$ and $ \ket{\Omega_0}$, lie in the same GNS Hilbert space.  
In such a case since the Petz reconstruction of the modes act on the same Hilbert space, it is interesting to compare them and find the relation between them.
Take the Petz map corresponding to two different GNS Hilbert space as 
$ \mathcal{R}^*_\Omega$ and $ \mathcal{R}^*_{\Omega_0}$. From \eqref{45}, we get
\begin{equation}\label{46}
    \begin{split}
        \langle \Tilde{a}_1 |\Delta_{bulk}^{1/2} |\Tilde{a}_2 \rangle =
        & \langle \mathcal{R}^*_{\Omega_0} (\Tilde{a}_1) |\Delta_{0}^{1/2} |
        \mathcal{R}^*_{\Omega_0}(\Tilde{a}_2) \rangle _{\Omega_0}
\\
        \langle \Tilde{a}_1 |\Delta_{bulk}^{1/2} | \Tilde{a}_2 \rangle =
        & \langle \mathcal{R}^*_{\Omega} (\Tilde{a}_1)
        |\Delta^{1/2} |\mathcal{R}^*_{\Omega}(\Tilde{a}_2) \rangle_{\Omega}.
    \end{split}
\end{equation}
Since the two are in the same GNS Hilbert space, their dual bulk geometry is the same and the left-hand sides of the equalities \eqref{46} coincide. By comparing the left-hand sides 
\begin{equation}
    \langle \Omega_0| \mathcal{R}^*_{\Omega_0} (\Tilde{a}_1^\dagger) \Delta_{0}^{1/2} 
        \mathcal{R}^*_{\Omega_0}(\Tilde{a}_2)| \Omega_0 \rangle
        = 
         \langle \Omega| \mathcal{R}^*_{\Omega} (\Tilde{a}_1^\dagger) \Delta^{1/2} 
        \mathcal{R}^*_{\Omega}(\Tilde{a}_2)| \Omega \rangle
\end{equation}
and using \eqref{47}, one reach to 
\begin{equation}\label{48}
    \mathcal{R}^*_\Omega (\Tilde{a}_\omega) = U~ \mathcal{R}^*_{\Omega_0} (\Tilde{a}_\omega) ~U^\dagger.
\end{equation}
Consider that we have the eternal black hole in the bulk and the vacuums $ \ket{\Omega_0}$ and $ \ket{\Omega}$ respectively correspond to the eternal black hole in equilibrium and have some excitations on the eternal black hole background created by $U$. Then, we get
\begin{equation}
    \mathcal{R}^*_{\Omega} (\Tilde{a}_{\omega,m}^\dagger) = \pi \big( \hat{O}_{\omega,m;L}\big)
    \qquad\qquad\qquad
        \mathcal{R}^*_{\Omega_0} (\Tilde{a}_{\omega,m}^\dagger) = \pi_0 \big( \hat{O}_{\omega,m;L}\big)
\end{equation}
where $ \pi_0$ and $ \pi$ are representations of the single trace operators on the GNS Hilbert spaces. From \eqref{48}, we reach to 
\begin{equation}\label{70}
     \pi \big( \hat{O}_{\omega,m;L}\big) = U~ \pi_0 \big( \hat{O}_{\omega,m;L}\big) ~U^\dagger.
\end{equation}

\section{Interior Petz reconstruction and Papadodimas-Raju proposal}\label{PR}

The idea of reconstructing the bulk modes in the left exterior via the Petz map  \eqref{50} is helpful even in the cases in which we have a one-sided black hole in the bulk. 
In this section, we attempt to construct the interior modes of a typical black hole microstate and we will see that we arrive to the same result as the Papadodimas-Raju proposal.

\subsection{Papadodimas-Raju proposal}

Consider a big one-sided black hole in AdS. In this case,  only the bulk modes outside the horizon can be described in the boundary theory using the HKLL  procedure, while describing the interior modes is much more challenging. For this purpose, a remarkable proposal has been introduced by Papadodimas and Raju in a series of papers \cite{papadodimas2013infalling, Papadodimas:2013jku,Papadodimas:2013wnh, Papadodimas:2015jra} 
to find a CFT description of the black hole interior when the system is in a pure state. Here we first shortly review the PR proposal. 
The main idea of the PR proposal is to focus on a code subspace of the CFT theory, which is created by acting with a \emph{small algebra}  on the corresponding pure state and then find the CFT description of the interior operator in a state-dependent manner just in the chosen code subspace. 

If the CFT pure state describes the stable black hole in AdS, it should be close to the thermal state. More precisely, we consider a typical pure state in the high-temperature phase of the gauge theory denoted by $ \ket{\Psi_0}$ (see Sec. \ref{3.4}).
The small algebra $ \mathcal{A}$ corresponds to simple experiments in the effective field theory in the bulk, i.e. the observables outside the horizon of the black hole. At large $N$ limit, $ \mathcal{A}$ can be thought of as the set of products of simple trace operators of low conformal dimensions up to $K$ number of these operators
\begin{equation}
    \mathcal{A} = \text{span}~ \{ O_{\omega_1},  O_{\omega_1} O_{\omega_2},...,  O_{\omega_1} O_{\omega_2}... O_{\omega_K} \}
\end{equation}
such that  $  O_{\omega_i} $ are the Fourier modes of the single trace operators and
$ \sum_i \omega_i \ll \mathcal{N}$, while $ \mathcal{N} $ is the CFT's central charge. Therefore, we do not have too many insertions and so $K\ll \mathcal{N}$.
Taking $ \mathcal{A}$ as the linear span of the products of the operators is equivalent to considering it as the set of all polynomials in the modes of the operators
\begin{equation}
    A_\alpha = \sum _{I} \alpha (I) ( O_{\omega_i})^{I(\omega_i)}
\end{equation}
where $ \alpha(I)$ are arbitrary coefficients and the sum runs over all functions $ I$. This set of polynomials forms a linear space. The size of the set of all such polynomials scales like $ \mathcal{N}^{K}$ and limit the dimension of this space $K$ should satisfy the constraint 
\begin{equation}
    \dim (\mathcal{A}) = \mathcal{N}^K \ll e^{\mathcal{N}}.
\end{equation}
Given a typical black hole microstate $ \ket{\Psi_0}$, one can define the code subspace, also called the small Hilbert space as
\begin{equation}
    \mathcal{H}_{\psi_0} = \text{span}~ \{ \mathcal{A} \ket{ \Psi_0}\}.
\end{equation}
A typical pure state for the observables in $ O \in \mathcal{A}$ can be approximated by the thermal state, i.e. 
\begin{equation}
    \bra{\Psi_0} O^\dagger O \ket{\Psi_0} = \frac{1}{Z} \tr ( e^{-\beta H} O^\dagger O )  + O(1/ \mathcal{N}).
\end{equation}

To describe interior modes in the dual CFT theory, the PR proposal requires doubling the set of operators $ \Tilde{O}_\omega$ corresponding to the operator in the small algebra which they call \emph{mirror operator}. These mirror operators commute with the original operators and moreover, they should be entangled with them in the pure state $\ket{\Psi_0}$ in an appropriate way 
to ensure they have the right properties in a given state of the CFT.
More concretely, the mirror operators defined as
\begin{equation}
\begin{split}
    \Tilde{O}_\omega \ket{\Psi_0} & = e^{-\beta \omega /2} O^\dagger_{\omega} \ket{ \Psi_0}
    \\
\Tilde{O}_\omega~  O_{\omega_1} ... O_{\omega_n}\ket{\Psi_0} &=   O_{\omega_1} ... O_{\omega_n}~\Tilde{O}_\omega \ket{\Psi_0} 
\end{split}
\end{equation}
Thus, demanding that the mirror operator has the correct behavior within low-point correlators 
in a given pure state leads to the set of linear equations for the mirror operators. 
As far as we do not have too many operator insertions, this set of operators can be solved in the full Hilbert space of the CFT.

\subsection{Black hole interior reconstruction using Petz map}

As we had in Sec. \ref{4.4}, we can start with just one black hole exterior. Having only access to the black hole exterior is equivalent to doing simple experiments on the boundary theory. In other words, the algebra of low-energy operators which we denote here as $ \mathcal{A}_{ex}$ is identified with the algebra of coarse-grained operators of the CFT, denoted by $ \mathcal{A}_{c}$ (see Sec. \ref{4.2})
\begin{equation}
    \mathcal{A}_{ex} = \mathcal{A}_{c}.
\end{equation}
From \eqref{19} and \eqref{20}, we know that for the black hole in equilibrium both 
$ \rho_{bulk}\big|_{\mathcal{A}_{ex}}$ and $ \rho_{CFT}\big|_{\mathcal{A}_{c}}$
are thermal and under the duality, we can identify them.

One can follow the discussion in appendix \ref{app} and build the GNS Hilbert spaces corresponding to these thermal states in the bulk and boundary over the algebras
$ \mathcal{A}_{ex}$ and $ \mathcal{A}_{c}$, we denote them as 
$\mathcal{H}^{GNS}_{ex}$ and $\mathcal{H}^{GNS}_{c}$
respectively. We take the vectors
\begin{equation}\label{51}
    \ket{\Omega_{ex}} \in \mathcal{H}^{GNS}_{ex} \qquad\qquad\qquad \ket{\Omega_{c}} \in \mathcal{H}^{GNS}_{c}
\end{equation}
as the cyclic vectors corresponding to 
$ \rho_{bulk}\big|_{\mathcal{A}_{ex}}$ and $ \rho_{CFT}\big|_{\mathcal{A}_{c}}$
which satisfy \eqref{A3}. Then, we have
\begin{equation}
\begin{split}\label{52}
     \mathcal{H}^{GNS}_{ex} &= \text{span} \{\mathcal{A}_{ex} \ket{\Omega_{ex}}\}
     \\
      \mathcal{H}^{GNS}_{c} &= \text{span} \{\mathcal{A}_{c} \ket{\Omega_{c}}\}.
\end{split}
\end{equation}
 The same as in Sec. \ref{4.4}, we will refer to the cyclic vectors in \eqref{51} as the GNS vacuums of the GNS Hilbert spaces \eqref{52}. We also identify the algebras with their representations on the GNS Hilbert spaces.

By accessing the information outside the black hole, the observable can not distinguish between all possible geometries of the entire bulk.
The bulk can be described as an eternal two-sided black hole or a one-sided black hole. If we know that in the bulk we have an eternal black hole that is dual to the TFD state on the boundary, we will reach exactly the setup that we discussed in Sec. \ref{4.4}.  
In this case, we have 
\begin{equation}
    \ket{\Omega_{ex}} = \ket{HH}, \qquad\quad
    \ket{\Omega_{c}} = \ket{\Omega_0}, \qquad\quad
     \mathcal{H}^{GNS}_{ex}=  \mathcal{H}^{(Fock)}_{\rho_{th}},\qquad\quad
     \mathcal{H}^{GNS}_{c}= \mathcal{H}_{TFD}
\end{equation}
and 
\begin{equation}
   ~~~ \mathcal{A}_{ex} = \mathcal{A}_{r,0}, \qquad\qquad
   ~ (\mathcal{A}_{ex})' = \mathcal{A}_{l,0} , \qquad\qquad
  \mathcal{A}_{c} = \mathcal{M}_{R} ,\qquad\qquad
    (\mathcal{A}_{c})' = \mathcal{M}_{L}.
\end{equation}
There is another possibility that we have a black hole microstate in equilibrium. The CFT dual of such a geometry is a typical state defined in \eqref{55}. We assume that the geometry corresponds to a typical state in the bulk has a smooth horizon and contains an interior region. 
Here, there does not exist a second copy of the  CFT as the left system and entire bulk is dual to just one CFT.

Consider a Cauchy slice $\Sigma$ in the bulk. We can divide it as 
$ \Sigma = \Sigma_{ex} \cup \Sigma_{in}$. 
$ \mathcal{A}_{ex} $ is the operator algebra of observable on $ \Sigma_{ex}$ and the same we can denote the algebra of operators on $ \Sigma_{in}$ as $ \mathcal{A}_{in}$. 
Since the two regions are spacelike separated, they commute with each other, and as they cover the entire Cauchy slice
\begin{equation}
    (\mathcal{A}_{ex})' = \mathcal{A}_{in}.
\end{equation}
Therefore, the commutant of the algebra $ \mathcal{A}_{ex}$ in the GNS Hilbert space 
$ \mathcal{H}^{GNS}_{ex}$ can be interpreted as the representation of the $ \mathcal{A}_{in}$
in the GNS Hilbert space. We identify them and denote the representation of the operator algebra inside the black hole in the GNS Hilbert space as  $ \mathcal{A}_{in}$ too. 
We can build $ \mathcal{A}_{in}$ in the GNS Hilbert space by conjugating with the modular conjugation as
\begin{equation}
    \mathcal{A}_{in} = J_{bulk} ~ \mathcal{A}_{ex} ~ J_{bulk}. 
\end{equation}
Hence, to each element of the algebra we associate an operator in the commutant as
\begin{equation}
     a_\omega \in\mathcal{A}_{ex}  \longrightarrow  \Tilde{a}_\omega \in \mathcal{A}_{in}
\end{equation}
where corresponds to the modes in the black hole interior.

On the other hand on the boundary, the degrees of freedom inside the black hole are encoded in the fine-grained observables of the CFT. In other words, the image of $ \mathcal{A}_{in}$ on the boundary, which we denote as $ \mathcal{A}_{in-bdy}$ is a subalgebra of fine-grained algebra of the CFT
\begin{equation}
 \mathcal{A}_{in-bdy} \subset  \mathcal{A}_{f} \subset \mathcal{L}(\mathcal{H}_{CFT}).
\end{equation}
Since $ \mathcal{A}_{in}$ is the commutant of $ \mathcal{A}_{ex}$, under the duality it should map to the commutant of $ \mathcal{A}_{c}$ on the $ \mathcal{H}^{GNS}_{c}$.
Therefore, we can identify the commutant of the algebra $ \mathcal{A}_{c}$ on the $ \mathcal{H}_{c}^{GNS}$ with the representation of the $ \mathcal{A}_{in-bdy}$ on the GNS Hilbert space
\begin{equation}
    (\mathcal{A}_{c})' = J_{bdy} ~ \mathcal{A}_{c} ~ J_{bdy}=\mathcal{A}_{in-bdy}
\end{equation}
where $ J_{bdy}$ is the modular conjugation for the vacuum $ \ket{\Omega_{c}}$ corresponding to the algebra $ \mathcal{A}_{c}$.
We also note that if the black hole is in the microstate $ \ket{\Psi_0}$, the GNS Hilbert space $ \mathcal{H}^{GNS}_{c}$
is isomorphic to the GNS Hilbert space obtained by acting the elements of $ \mathcal{A}_{c}$ on the state $ \ket{\Psi_0}$. 

Following the discussion in Sec. \ref{4.3}, we can map the algebra of the operator outside the horizon $ \mathcal{A}_{ext}$ to the coarse-grained algebra $ \mathcal{A}_{cg}$ of the boundary theory through the Petz map in \eqref{56} while the Schwarzschild modes mapped as \eqref{57}.
But here instead we do not know the isometry that maps the interior modes to the boundary theory unlike in the case of the eternal black holes where the left exterior can also be mapped to the left CFT via the second copy of the HKLL map.
Moreover, we do not even know any global mapping like the global HKLL map in the pure AdS spacetime that maps the entire bulk to the entire boundary for us to use the same logic as the one that has been done to find the Petz map in order to reconstruct the entanglement wedge of a given boundary region. 

The discussion we had in the previous section suggests the idea that we can use the definition of the Petz map in modular theory and map the interior modes to the boundary via \eqref{50}. In addition to that from the bulk perspective for a black hole in a typical microstate the geometry locally is the same as eternal black holes, and even for the late time bulk correlation functions between the operators inside and outside the horizon of the collapsing star geometry, it is known that they can be approximated very well by the correlators of the operators in regions \RNum{1} and \RNum{2} of an eternal black hole \cite{papadodimas2013infalling}. Thus indeed here the interior modes play the role of the modes coming from the left side of the eternal geometry. But the important difference is that the commutant of the image of the operator algebra in the region \RNum{1} is no longer in the second CFT but rather it represents a subalgebra of fine-grained operators in the original CFT.  

As a consequence of all the discussions above, we introduce the Petz map that encodes the interior modes of a black hole microstate in the dual CFT $ \mathcal{R}^*_{c,in} : \mathcal{A}_{in} \rightarrow \mathcal{A}_{in-bdr}$ as 
\begin{equation}
    \mathcal{R}^*_{c,in} (.) = \mathcal{J}_{bdy} \circ \mathcal{R}'_{c,\Omega} \circ \mathcal{J}_{bulk} (.)
\end{equation}
which leads to
\begin{equation}\label{60}
    \begin{split}
        \Tilde{a}_{\omega,m} & \longrightarrow 
         \mathcal{R}^*_{c,in}(\Tilde{a}_{\omega,m}) =  J_{bdy} ~\hat{O}_{\omega,m;c}~ J_{bdy} \equiv \Tilde{O}_{\omega,m}\in \mathcal{L}(\mathcal{H}^{GNS}_{c})
        \\
     \Tilde{a}^\dagger_{\omega,m} & \longrightarrow 
         \mathcal{R}^*_{c,in}(\Tilde{a}^\dagger_{\omega,m}) =  J_{bdy} ~\hat{O}^\dagger_{\omega,m;c}~ J_{bdy} \equiv \Tilde{O}_{\omega,m} ^\dagger\in \mathcal{L}(\mathcal{H}^{GNS}_{c}).
    \end{split}
\end{equation}
\begin{figure}[h]
    \centering
    \begin{tikzpicture}[>=stealth]
    
    
    \node at ( 4.7,0) (L) {$\mathcal{A}_{in-bdy}$};
    \node at ( 5,3) (3) {$\mathcal{A}_{in}$};
    
    \node at ( 9,0)  (R){$\mathcal{A}_{c}$};
    \node at ( 9,3)  (1){$\mathcal{A}_{ex}$};
    \draw [->, dashed] (9,2.6) -- (9,0.3);
    \node at ( 9.9 ,1.5)  {$\mathcal{R}'_{c-\Omega}$};
    
    \draw [->] (R.west) -- (L.east);
    \draw [->] (3.east) -- (1.west);
    
    \node at ( 7,3.5)  {$\mathcal{J}_{bulk}$};
    \node at ( 7,-0.5)  {$\mathcal{J}_{bdy}$};
  
    \end{tikzpicture}
\end{figure}

We can also find the dual operator through its insertion between the vectors in the GNS Hilbert space. From \eqref{52}, it is clear that every vector  $\ket{a} \in \mathcal{H}^{GNS}_{c}$ can be obtained by the action of an element of the algebra $ a \in \mathcal{A}_{c}$ on the GNS vacuum
\begin{equation}
    \ket{a} \equiv \pi(a) \ket{\Omega_{c}}
\end{equation}
where $ \pi (a)$ is the representation of $ a$ in the GNS Hilbert space. Then, we have
\begin{equation}
\begin{split}
        \bra{a} \Tilde{O}_{\omega,m} \ket{b} &= \bra{\Omega_{c}} \pi (a)^\dagger ~\Tilde{O}_{\omega,m} ~\pi(b) \ket{\Omega_{c}}
        = \bra{\Omega_{c}} \pi (a)^\dagger ~\pi(b) ~\Tilde{O}_{\omega,m} \ket{\Omega_{c}}
        \\
     \bra{a} \Tilde{O}^\dagger_{\omega,m} \ket{b} &= \bra{\Omega_{c}} \pi (a)^\dagger~ \Tilde{O}^\dagger_{\omega,m} ~\pi(b) \ket{\Omega_{c}}
     = \bra{\Omega_{c}} \pi (a)^\dagger~\pi(b) ~ \Tilde{O}^\dagger_{\omega,m} \ket{\Omega_{c}}.
\end{split}
\end{equation}
To go ahead, we remind that when the black hole is  in equilibrium from \eqref{19} we have 
\begin{equation}\label{59}
    \ket{\Omega_{c}} \Big|_{\mathcal{A}_{c}} = \rho_{th,c}= \frac{1}{Z^{c}_\beta} e^{-\beta H_{c}},
\end{equation}
One can obtain that for every $ a \in \mathcal{A}_{cg}$
\begin{equation}
    J_{bdy}~ a~ J_{bdy} \ket{\Omega_{cg}} = \rho_{th,c}^{1/2} ~a^\dagger~ \rho_{th,c}^{-1/2} \ket{\Omega_{c}} = e^{-\beta H_{c}/2} ~a^\dagger~ e^{\beta H_{c}/2} \ket{\Omega_{c}}.
\end{equation}
Considering \eqref{59} roughly speaking, we can also interpret the vacuum $ \ket{\Omega_{c}}$ as a TFD state with respect to  $ H_c$ in the GNS Hilbert space as 
\begin{equation}
    \ket{\Omega_{c}} = \sum _i e^{-\beta E_i /2} \ket{E_i}_c \ket{E_i}_f
\end{equation}
where $ E_i$s are the energy eigenvalues of the coarse-grained Hamiltonian.
As it is mentioned, the coarse-grained observables are the GFFs around the thermal state of the CFT. Thus, the coarse-grained Hamiltonian should be in the form of 
\begin{equation}
   H_c = \sum_{\omega,m} \omega ~O_{\omega,m;c}^\dagger O_{\omega,m;c}
\end{equation}
where the operators $O_{\omega,m;c}$, the projection of the Fourier modes of the single trace operators onto the coarse-grained part of the system, can be identified with the representation of the single trace operators in the GNS Hilbert space $ \mathcal{H}^{GNS}_{c}$.
They satisfy 
\begin{equation}
    [H_c, O_{\omega,m;c}] = - \omega ~O_{\omega,m;c} \qquad\qquad\quad
    [H_c, O^\dagger_{\omega,m;c}] = \omega ~O^\dagger_{\omega,m;c} 
\end{equation}
and therefore, one can obtain that 
\begin{equation}\label{61}
    \begin{split}
        e^{-\beta H_{c}/2} &~O_{\omega,m;c}~ e^{\beta H_{c}/2} = e^{\beta \omega /2} ~O_{\omega,m;c}
\\
        e^{-\beta H_{c}/2} &~O_{\omega,m;c}^\dagger~ e^{\beta H_{c}/2}
        = e^{-\beta \omega /2}~ O_{\omega,m;c}^\dagger.
    \end{split}
\end{equation}
In the end, using the definition of the operator $ \Tilde{O}_{\omega,m}$ \eqref{60} and the relations \eqref{61}, we reach to  
\begin{equation}
\begin{split}
        \bra{a} \Tilde{O}_{\omega,m} \ket{b} &
        =  e^{-\beta \omega /2}~ \bra{\Omega_{c}} \pi (a)^\dagger ~\pi(b) ~ O_{\omega,m;c}^\dagger\ket{\Omega_{c}} =  e^{-\beta \omega /2}~ \tr (\rho_{th,c}~ a^\dagger b ~O^\dagger_{\omega,m}) + O(1/ \mathcal{N})
        \\
     \bra{a} \Tilde{O}^\dagger_{\omega,m} \ket{b} &= e^{\beta \omega /2}~\bra{\Omega_{c}} \pi (a)^\dagger~\pi(b) ~ O_{\omega,m;c} \ket{\Omega_{c}} =e^{\beta \omega /2}~ \tr ( \rho_{th,c}~ a^\dagger b ~O_{\omega,m}) + O(1/ \mathcal{N}).
\end{split}
\end{equation}

Moreover, we also obtain 
\begin{equation}
\begin{split}
    \Tilde{O}_{\omega,m} \ket{\Omega_c} & = e^{-\beta \omega /2}~ O^\dagger_{\omega,m;c} \ket{ \Omega_c}
    \\
\Tilde{O}_{\omega,m} ~  O_{\omega_1,m_1;c}~ ...~ O_{\omega_n,m_n;c}\ket{\Omega_{c}} &= O_{\omega_1,m_1;c}~ ...~ O_{\omega_n,m_n;c}~ \Tilde{O}_{\omega,m} \ket{\Omega_c}
\end{split}
\end{equation}
which is equivalent to the operators result in the PR proposal but it is obtained in a more concrete way. The black hole microstate is replaced by the GNS vacuum state $ \ket{\Omega_c}$ and considering the code subspace in the PR proposal is equivalent to work in the GNS Hilbert space.

The GNS Hilbert space $ \mathcal{H}^{GNS}_{c}$ can be also constructed through the action of $ \mathcal{A}_c$ on other cyclic and separating vectors $\ket{\Omega_c'}$ belong to $ \mathcal{H}^{GNS}_{c}$. They can be related to  $ \ket{\Omega_c}$ via a unitary operator $ u \in \mathcal{L}(\mathcal{H}^{GNS}_{c})$ as
\begin{equation}\label{71}
    \ket{\Omega_c'} = U \ket{\Omega_c}
\end{equation}
and the vectors in the GNS Hilbert space can be identified as 
\begin{equation}
    \mathcal{H}^{GNS}_{c} = \text{span} \big\{ | a'\rangle \equiv \pi (a) |\Omega_c'\rangle ~| ~\forall a \in \mathcal{A}_c \big\}.
\end{equation}

If we denote the representation of the operator dual to the interior mode $ \Tilde{a}_{\omega,m}$ in the GNS Hilbert space is built over the vector $ \ket{\Omega_c'}$ by $ \Tilde{O}_{\omega,m}'$, from \eqref{70} we reach to  
\begin{equation}
    \Tilde{O}_{\omega,m}' = U ~ \Tilde{O}_{\omega,m}~U^\dagger
\end{equation}
while $ \Tilde{O}_{\omega,m}$ is the representation of the dual operator we defined in the GNS Hilbert space over the thermal state \eqref{60}.
In particular, for the matrix elements of the operator, we find that 
\begin{equation}
\begin{split}
     \bra{a'} \Tilde{O}_{\omega,m}' \ket{b'} &= \bra{\Omega_{c}'} \pi (a)^\dagger ~\Tilde{O}_{\omega,m}' ~\pi(b) \ket{\Omega_{c}'}
     \\
     & =  \bra{\Omega_{c}} U^\dagger \pi (a)^\dagger U~\Tilde{O}_{\omega,m}~U^\dagger\pi(b) U\ket{\Omega_{c}}.
     \end{split}
\end{equation}
and to be compatible with the PR proposal, we can also find that
\begin{equation}
    \begin{split}
        \Tilde{O}_{\omega,m}' \ket{\Omega_c'} & = U ~\Tilde{O}_{\omega,m} \ket{\Omega_c}
        \\
        &= U~ \rho_{th,c}^{1/2} ~ O_{\omega,m;c}^\dagger~ \rho_{th,c}^{-1/2} U^\dagger\ket{\Omega_{c}'}
        \\
        &=  U~  e^{-\beta\omega/2}~ O_{\omega,m;c}^\dagger~  U^\dagger\ket{\Omega_{c}'}.
    \end{split}
\end{equation}

The vectors in \eqref{71} correspond to an equilibrium black hole background which is excited by some sources.
It can be provided by turning on a source for some CFT operators. The unitary operator in \eqref{71} is indeed the representation of the composite operator which creates that excitation.
The simplest cases for the unitary operator $U$ that result in a cyclic and separating vector $ \ket{\Omega_c'}$ correspond to the local unitaries as 
\begin{equation}\label{72}
    V_c \in \mathcal{A}_c
\end{equation}
related to the excitation only on the region \RNum{1}, or 
\begin{equation}\label{73}
    W_{f} \in \mathcal{A}_c'= \mathcal{A}_{in-bdy}.
\end{equation}
correspond to the excitation behind the black hole horizon.
If we have the unitaries as in \eqref{72}, we get 
\begin{equation}
    \langle a'| ~\Tilde{O}_{\omega,m}' ~| b'\rangle = \langle~ V_c^\dagger ~ a~ V_c |~ \Tilde{O}_{\omega,m}~ |V_c^\dagger ~ b~ V_c ~\rangle
\end{equation}
since $ V_c^\dagger ~ a~ V_c \in \mathcal{A}_c$, and in the case that we have some excitation inside the black hole, i.e. act with some unitary in the form of \eqref{73}, we get
\begin{equation}
     \langle a'| ~\Tilde{O}_{\omega,m}' ~| b'\rangle = \langle a|~\Tilde{O}_{\omega,m}~ |b\rangle
\end{equation}
as $ W_f$ and $ W_f^\dagger$ commute with every $ a,~b \in \mathcal{A}_c$.
As a result, as it is expected we see that if we access only the coarse-grained observables, we can not detect what happening behind the black hole horizon. 
It is good to mention here that acting with \eqref{72} corresponds with considering the near-equilibrium states in the PR proposal and like in the thermal state, the Petz map reconstruction will reach to exactly the same result as the PR proposal.

\section{Discussion}

In order to study the evaporating black hole in AdS, one can use absorbing boundary conditions. In \cite{penington2020entanglement, almheiri2019entropy}, it has been shown that exactly at Page time, there is a phase transition in the location of the quantum extremal surface. The new Ryu-Takayanagi surface lies slightly inside the black hole event horizon. Thus after Page time, some parts of the interior are now encoded in the early Hawking radiation, or in other words, it can be reconstructed through the bath. 

In order to study the reconstruction of the interior, let us first consider a general entangled system. The CFT can be entangled with another CFT or a collection of qubits. we refer to another system as a bath. Here, we consider the entangled state as 
\begin{equation}
    \ket{\Psi_{en}} = \sum _i \alpha_i \ket{\psi_i} \otimes \ket{\Tilde{i}}
\end{equation}
where $ \alpha_i$ are some coefficients, $ \ket{\psi_i}$ are orthonormal states in the original CFT, and $ \ket{\Tilde{i}}$ are states in the bath. 
The sum can be over a small number of states or an exponentially large number.

We denote the coarse-grained algebra of the original CFT as $\mathcal{A}_{cg}$ and the operator algebra of the bath as $\mathcal{B}$. We define 
\begin{equation}
    \mathcal{A}_{bdy} = \mathcal{A}_{cg} \otimes \mathcal{B}.
\end{equation}
The code subspace here is the set of states obtained by acting the algebra $ \mathcal{A}_{bdy}$ over the state $ \ket{\Psi_{en}}$. The corresponding subspace has the structure of a Hilbert space that can be made via GNS construction 
\begin{equation}
    \mathcal{H}^{GNS}_{en}\cong \mathcal{A}_{bdy} \ket{\Psi_{en}}. 
\end{equation}
This set of states generally is bigger than $ \mathcal{A}_{cg} \ket{\Psi_{en}}$ and in some specific cases like when the entangled state is TFD state, these two sets coincide.

In general, as it was discussed in \cite{Papadodimas:2015jra}, the GNS Hilbert space can be decomposed into the direct sum of $ \mathcal{H}^j_{\Psi_{en}}$ while all are closed under the action of the coarse-grained algebra
\begin{equation}
    \mathcal{H}^{GNS}_{en} = \oplus _j \mathcal{H}^j_{\psi_{en}}.
\end{equation}
For each $j$, one can identify a unique state $ \ket{\psi_{en}^j} \in \mathcal{H}^j_{\psi_{en}}$ which is an equilibrium state with respect to $ \mathcal{A}_{cg}$
\begin{equation}
  \ket{\psi_{en}^j} \big|_{\mathcal{A}_{cg}} = \rho _{th} ^{c}  
\end{equation}
and entire $ \mathcal{H}^j_{\psi_{en}}$ can be generated by acting with $ \mathcal{A}_{cg}$ on $\ket{\psi_{en}^j} $.

For the exterior of the black hole, the same as \eqref{57}, we have the mapping $ \mathcal{R}^*_{ext}: \mathcal{A}_{ext} \rightarrow \mathcal{A}_{cg}$ and from its $ \rho$-dual, we can find the Petz map from the operator algebra of the interior to the commutant of the representation of the coarse-grained algebra on the boundary
\begin{equation}
    \mathcal{R}^*_{in}: \mathcal{A}_{in} \rightarrow \mathcal{M}'_{cg}.
\end{equation}
The interior part of a Cauchy slice at late time can be divided into the island and the remaining part of the interior which can be reconstructed from the original CFT. Thus we consider $ \mathcal{A}_{in} = \mathcal{A}_{island} \otimes \mathcal{A}_{in-CFT}$.
The algebra $ \mathcal{M}'_{cg}$ is the representation of the $ \mathcal{A}'_{cg}= \mathcal{B}\otimes\mathcal{A}_{fg}$ while $ \mathcal{A}_{fg}$
is the fine-grain algebra of the original CFT.

In the GNS Hilbert space $ \mathcal{R}^*_{in}$ can be obtained from the direct sum of the mapping in each $ \mathcal{H}_{en}^j$ as 
\begin{equation}
    \mathcal{R}^*_{in} = \oplus_j \mathcal{R}^*_{in,j}
\end{equation}
Each $ \mathcal{R}^*_{in,j}$ can be obtained from the same approach as obtaining the Petz dual map in modular theory. In each mapping $ \mathcal{R}^*_{in,j}$ depending on the structure of the entanglement in $ \ket{\psi^j_{en}}$, the interior can be map to the commutant of the $ \mathcal{A}_{cg}$ in the $ \mathcal{H}_{en}^j$ that can be the representation of a subalgebra of the fine-grained algebra or the algebra of the bath system.

From the island conjecture, it is expected that 
\begin{equation}
    \begin{split}
         \mathcal{R}^*_{in}(a)& = \oplus_j \mathcal{R}^*_{in,j}(a) \in \mathcal{A}_{fg} \qquad \forall a \in \mathcal{A}_{in,CFT}
         \\
         \mathcal{R}^*_{in}(a)& = \oplus_j \mathcal{R}^*_{in,j}(a) \in \mathcal{B}~~~ \qquad \forall a \in \mathcal{A}_{island}
    \end{split}
\end{equation}

Up to this point, there is not exist any microscopic proof of the island conjecture in the literature.
Doing the exact calculation of the Petz map reconstruction of an evaporating black hole can be a good check of the island conjecture.



\acknowledgments
I would like to thank M.Mirbabayi and K. Papadodimas for useful
discussions and comments on the draft. In particular, I would like to thank my supervisor, K. Papadodimas for
initial discussions on a possible connection of the Petz map
reconstruction of the black hole interior with the Papadodimas-Raju
proposal. I would like to thank
particularly M. Bertolini and M. Serone for their invaluable support
during this work.
The research is partially supported by INFN Iniziativa Specifica - String Theory and Fundamental Interactions project.

\appendix

\section{Tomita-Takesaki theory in a nutshell}\label{app}

In this section, we briefly review the Tomita-Takesaki theory. It is mostly based on \cite{haag2012local, furuya2020real, lashkari2021modular}.

The set of all bounded, linear operators acting on a Hilbert space $\mathcal{H}$ is denoted by $ \mathcal{L}(\mathcal{H})$. A subset $ \mathcal{A}\subset\mathcal{L}(\mathcal{H}) $ which is closed under Hermitian conjugation, addition, multiplication, and  closed under the weak convergent limit that also contains the unit operator is called a \emph{von Neumann} algebra. 
For a given $ \mathcal{A}$, the set of all bounded operators which commute with every elements of $\mathcal{A}$ is called the \emph{commutant} of $\mathcal{A}$
\begin{equation}
    \mathcal{A}' = \{b \in \mathcal{L}(\mathcal{H}) | ab = ba, \forall a \in \mathcal{A}\}
\end{equation}
which itself is a von Neumann algebra. For any von Neumann algebra $ \mathcal{A}$ on $ \mathcal{H}$, we have $ \mathcal{A}'' = (\mathcal{A}')' = \mathcal{A}$.
Another von Neumann algebra which is induced by $ \mathcal{A}$ is the \emph{center} of the algebra, denoted by $ Z_{\mathcal{A}}= \mathcal{A} \cap \mathcal{A}'$. 

A representation of the algebra $ \mathcal{A}$ in a Hilbert space $\mathcal{H}$ is a map $\pi$ from the algebra to the bounded operators on $\mathcal{H}$ such that $ \pi(ab) = \pi (a) \pi (b)$ and $ \pi (a^*) = \pi (a) ^\dagger$. The map $ \pi $ is unital if $ \pi (I)= I $. 
A linear form over $\mathcal{A}$ is a function from algebra to the complex numbers $ \phi :\mathcal{A} \rightarrow \mathbb{C}$ such that 
\begin{equation*}
    \phi (\alpha a + \beta b) = \alpha \phi (a) + \beta \phi (b)  \qquad \forall a,b \in \mathcal{A} , \alpha , \beta \in \mathbb{C}.
\end{equation*}
It is called positive if $ \phi (aa^*) \geq 0, \forall a \in \mathcal{A}$, and normalized if $ \phi (I) = 1$. A normalized, positive linear form is called a \emph{state} on a von Neumann algebra. 

Following the GNS construction, for each positive linear form 
$\phi$ over $ \mathcal{A}$, one can build a Hilbert space $ \mathcal{H}_{\phi}$ and a representation $ \pi _ \phi$ of the algebra $ \mathcal{A}$ by linear operators acting on $ \mathcal{H}_\phi$. The state $ \phi$ defines a Hermitian scalar product on $ \mathcal{A}$ as 
\begin{equation}
    \langle a | b \rangle = \phi (a^* b) \qquad \forall a,b \in \mathcal{A}.
\end{equation}
A vector $ \ket{\Psi} \in \mathcal{H}$ is called $cyclic$ for an algebra $ \mathcal{A}$ if the set of $ a \ket{\Psi}$ for $ a \in \mathcal{A} $ are dense in $ \mathcal{H}$ and $separating$ if the condition $ a\ket{\Psi} =0$ implies that $ a=0$. If $ \ket{\Psi}$ is cyclic and separating for $\mathcal{A}$, it is also for $\mathcal{A}'$. And naturally, a representation $ \pi$ is called cyclic if there exists a vector $ \ket{\Psi}$ in the representation space $\mathcal{H}$ such that $ \pi (\mathcal{A}) \ket{\Psi}$ is dense in $\mathcal{H}$. The GNS construction provides a cyclic representation with the cyclic vector $\ket{\Psi}$ which 
\begin{equation}\label{A3}
  \psi (a)= \langle \Psi | \pi _\psi (a)| \Psi \rangle  
\end{equation}
that is familiar form of the expectation values in quantum mechanics. For the faithful linear form, we will identify $ \mathcal{A}$ with $ \pi _ \psi (\mathcal{A})$.

Moreover, there is a correspondence between superoperators on $ \mathcal{A}$ and linear operators acting on $ \mathcal{H}$.  A linear map from algebra to itself $ \mathcal{T}: \mathcal{A} \rightarrow \mathcal{A} $ is called a $superoperator$. A superoperator is called unital if $ \mathcal{T}(I)=I$ and $ \phi$-preserving  if 
$ \phi (\mathcal{T} (a))= \phi (a)$ for all $ a \in \mathcal{A}$. For a generic von Neumann algebra, every normal superopeartor has a corresponding operator in the GNS Hilbert space. However, the converse does not always hold, like the local algebra of QFT. Although, in matrix algebra, the correspondence is one-to-one. 
The GNS Hilbert space operator $ T_\psi \in \mathcal{L}(\mathcal{H}_\psi)$   corresponding to the superoperator $\mathcal{T}$ is defined in such a way that
\begin{equation}
    T_\psi (a\ket{\Psi}) = \mathcal{T}(a)\ket{\Psi}
\end{equation}
for all $ a\in \mathcal{A}$.
 If $ \mathcal{T}$ is unital, $ T_\psi$ leaves $ \ket{\Psi}$ invariant and if $\mathcal{T}$ is $\psi$-preserving, $ T^\dagger _\psi$ also leaves $ \ket{\Psi}$ invariant.

Before proceeding, to have more intuition, let us first consider the Type \RNum{1} von Neuman algebra, $i.e.$ the algebra of $ d \times d $ complex matrices acting irreducibly on the Hilbert space $ \mathcal{K}$ of a $d$-level system, denoted by $\mathcal{L}(\mathcal{K})$. In such a system, states are described by a positive, semi-definite, Hermitian operator of trace one $ \rho \in \mathcal{L}(\mathcal{K})$, which is called $density$ $matrix$.
The set of all density matrices on $ \mathcal{K}$ denoted by $ \mathcal{S}(\mathcal{K})$.
Corresponding to the state $ \rho$, one can define a map $ \phi _\rho : \mathcal{L}(\mathcal{K}) \rightarrow \mathbb{C}$ given by $ \phi _ \rho (a)= \tr (\rho a)$, such that for any observable on the system gives us its expectation value on the state $ \rho$. 

Given a density matrix 
\begin{equation}
  \rho = \sum _i \lambda _i ^ 2 \ket{i}\bra{i}  
\end{equation}
We can always purify the state by coupling it with a second system with the Hilbert space $ \mathcal{K}'$ such that $ \text{dim} ~\mathcal{K}' \geq \text{rank}~ \rho $. The Schmidt decomposition always guarantees that there exists such a system that equality holds. Here, we take $ \mathcal{K}'  = \text{span} \{ \ket{i'}\}$ to be isomorphic to $ \mathcal{K}$. 
The state $ \rho $ on $ \mathcal{K}$ can be purified by a vector 
\begin{equation}
   |\rho^{1/2}\rangle = \sum _ i \lambda _i \ket{i} \ket{i'} \in \mathcal{K} \otimes \mathcal{K}' 
\end{equation}
such that $ \rho = \tr _{\mathcal{K}'} |\rho^{1/2}\rangle \langle\rho^{1/2}|$.
Therefore, we can consider $ \mathcal{A} = \mathcal{L}(\mathcal{H}) \otimes I_{\mathcal{K}'} $  as 
a von Neumann algebra on $\mathcal{K} \otimes\mathcal{K}'= \mathcal{H}_ \rho  $. Here, the commutant is simply $  \mathcal{A}' =I_{\mathcal{K}} \otimes \mathcal{L}(\mathcal{K}')  $ (Sometimes, for simplicity, we just refer them as  $\mathcal{A} = \mathcal{L}(\mathcal{K}) $ and $ \mathcal{A}' = \mathcal{L}(\mathcal{K}')$). 
 The vector $ |\rho^{1/2}\rangle$ is cyclic and separating for the algebra $ \mathcal{A}$ if and only if  $ \rho$ is full-rank. We have also 
 \begin{equation}
     \phi _ \rho (a)= \tr (\rho a) = \langle \rho ^{1/2} |(a \otimes I )| \rho ^{1/2}\rangle
 \end{equation}
 for all $a \in \mathcal{A}$. Thus, following the GNS construction, we can find a cyclic representation of $ \mathcal{A}$ with the cyclic vector $|\rho^{1/2}\rangle$. The map from 
  $ \mathcal{A} \rightarrow \mathcal{H}_ \rho$ defined as 
  \begin{equation}
      a \longrightarrow \ket{a}_ \rho = \big(a \otimes I \big)|\rho^{1/2}\rangle=\sum _ i \lambda _i \big(a\ket{i}\big) \ket{i'},
  \end{equation}
and $ \mathcal{H}_ \rho$ is nothing but the set of vectors $ \big(a \otimes I \big)| \rho ^{1/2}\rangle$ endowed  with the inner product 
\begin{equation}\label{27}
    \langle a | b \rangle _ \rho =\phi _ \rho (a^ \dagger b)= tr \big(\rho a^ \dagger b\big) = \langle \rho^{1/2} |\big(a^\dagger b \otimes I \big)| \rho^{1/2} \rangle.
\end{equation}
In addition, for every $ a\in \mathcal{A}$, there exists an operator $ a_m' \in \mathcal{A}'$ that creates the same vector in the $ \mathcal{H}_\rho$ as $a$
\begin{equation}
    (a \otimes I)\ket{\rho } =(I \otimes a_m')\ket{\rho }
\end{equation}
which is called the $mirror$ operator of $a$ and it is given by
\begin{equation}
    a_m' = \rho ^{1/2} a^T \rho ^{-1/2},
\end{equation}
where transpose is taken in the $ \rho$ eigenbasis.

More generally, one can consider a generic algebra $ \mathcal{A}$ on some Hilbert space $ \mathcal{H}$, and for every cyclic and separating state $ \ket{\Psi}$ for $ \mathcal{A}$ on $ \mathcal{H}$ creates a GNS Hilbert space. 
There is a classification of von Neumann algebras on finite-dimensional Hilbert space corresponding to the center of the algebra.
One special case is when the center is trivial $ \mathcal{Z}_{\mathcal{A}} = \{ \lambda I \}$. 
In such a case the algebra is called a $factor$. If $ \mathcal{A}$ is a factor on $ \mathcal{H}$, there always exists a tensor factorization of Hilbert space $ \mathcal{H} = \mathcal{H}_A \otimes \mathcal{H}_{\Bar{A}}$ such that $ \mathcal{A}$ is just the set of all linear operators on one tensor factor $ \mathcal{A}= \mathcal{L}(\mathcal{H}_A) \otimes I _{\Bar{A}}$.
For a generic case that $ \mathcal{A}$ is not a factor, there is a decomposition of the Hilbert space as $ \mathcal{H} = \oplus _\alpha ( \mathcal{H}_ {\mathcal{A} _ \alpha } \otimes \mathcal{H}_ {{\Bar{A}}_\alpha})$ which $\mathcal{A}$ is block-diagonal $ \mathcal{A} = \oplus _\alpha \big( \mathcal{L}(\mathcal{H}_ {\mathcal{A} _ \alpha }) \otimes I_ {{\Bar{A}}_\alpha}\big)$ .

Here, the set of states on the algebra $ \mathcal{A}$ is the intersection of the algebra with the set of states on the Hilbert space $ \mathcal{S}(\mathcal{A})= \mathcal{A} \cap \mathcal{S}(\mathcal{H})$.
Any state $ \rho \in \mathcal{S} (\mathcal{A})$ is connected with the standard definition of state on von Neumann algebra by linear form $ \phi _\rho (a) = tr(\rho a)$ for all $ a \in \mathcal{A}$.
Moreover, for any state $ \rho$ on $ \mathcal{H}$, there exists a $unique$ restriction $ \rho \big|_{\mathcal{A}}$ on $ \mathcal{S}(\mathcal{A})$ such that 
\begin{equation}
    \phi _ {\rho |_ {\mathcal{A}}}(a) = \phi_ {\rho} (a)\qquad \forall a \in \mathcal{A}.
\end{equation}

For now, consider a factor $ \mathcal{A} = \mathcal{L} (\mathcal{H}_A) \otimes I_ {\Bar{A}} $ on $ \mathcal{H} = \mathcal{H}_A \otimes \mathcal{H}_{\Bar{A}} $ and a state $ \rho$ on $ \mathcal{H}$.  The restricton of any $ \rho \in \mathcal{S}(\mathcal{H})$ on $ \mathcal{A}$ is 
\begin{equation}
    \rho \big| _ {\mathcal{A}} = \rho _ A
\otimes \frac{1}{ |\Bar{A}|}I_{\Bar{A}},
\end{equation}
that $ \rho _ A \equiv \tr _{\Bar{A}} \rho$ is the reduced density matrix of the subsystem $A$ and $ |\Bar{A}|= \dim (\mathcal{H}_ {\Bar{A}})$.
One can follow the discussion above for $ \rho _ A$ and create the GNS Hilbert space representation of $ \mathcal{A} = \mathcal{L}(\mathcal{H}_A)$ as $ \mathcal{H}_ {\rho _A} = \mathcal{A} ~|\rho _A ^{1/2} \rangle$. If $ \rho _A$ is full-rank and the two tensor factors have the same dimensionality, this GNS Hilbert space is isomorphic to the original $ \mathcal{H}$. Otherwise, it is isomorphic to one subspace of $\mathcal{H}$.  

Let us look at some important superoperators and their corresponding operators in the GNS Hilbert space:

An important anti-linear superoperator which defines complex conjugation is the modular map $ \mathcal{S}(a) = a^\dagger$. Its GNS Hilbert space operator correspondence, called $Tomita$ operator $ S_\rho: \mathcal{H}_\rho \rightarrow \mathcal{H}_\rho$, acts as
\begin{equation}
    S_\rho \big(a \otimes I\big)|\rho^{1/2}\rangle  = \big(a^ \dagger \otimes I\big) |\rho ^{1/2} \rangle.
\end{equation}
It is clear that $ S_\rho^2 = I$, thus $ S_\rho$ is invertible. We also have $ S_\rho |\rho ^{1/2} \rangle = |\rho ^{1/2} \rangle $. As $ S_\rho$ is anti-linear, the $ S^\dagger _ \rho$ is defined by 
\begin{equation}
    \bra{a}S_\rho^\dagger~ b \rangle _\rho =  \bra{S_\rho ~ a}b \rangle ^* _\rho =  \bra{b}S_\rho ~a \rangle _\rho
    \end{equation}
for all $ a, b \in \mathcal{A}$. The Tomita operator for the commutant $ \mathcal{A}'$ is $ S_\rho ' = S _\rho ^ \dagger$.

Another important anti-linear superoperator is the  one-to-one corresponding map $ \mathcal{J}_ \rho : \mathcal{A} \rightarrow \mathcal{A}'$ between operators in $\mathcal{A} $ and $ \mathcal{A}'$, such that
\begin{equation}
    \mathcal{J}_\rho (\ket{i}\bra{j}) = \ket{i'}\bra{j'}.
\end{equation}
The  operator corresponding to it is the anti-linear map $ J_\rho :\mathcal{H}_\rho \rightarrow \mathcal{H}_\rho $ called $ modular$  $conjugation$ that acts on the GNS Hilbert space as
\begin{equation}\label{1}
     J_\rho \big(a \otimes I\big)|\rho ^{1/2} \rangle = \big(I \otimes (a^ \dagger)^T\big)|\rho ^{1/2} \rangle
\end{equation}
where the transpose is in the $\rho$ eigenbasis. In other words in this basis, $ J_\rho$ acts as 
\begin{equation}\label{2}
    J_\rho c_i \ket{i}\ket{j'} = c_i^* \ket{j} \ket{i'}.
\end{equation}
It also leaves $|\rho ^{1/2} \rangle$ invariant.

We also have the $ relative$ $ modular$ operator $\mathcal{D}_{\sigma| \rho} : \mathcal{A} \rightarrow \mathcal{A} $ as an superoperator on $ \mathcal{A}$ defined as $\mathcal{D}_{\sigma| \rho } (a) = \sigma a \rho ^{-1}$, where $ \rho$ and $ \sigma$ are two full-rank density matrices. Its corresponding operator on the GNS Hilbert space is $\Delta_{\sigma| \rho} : \mathcal{H}_\rho \rightarrow \mathcal{H}_\rho $. Since by definition, we have
\begin{equation}
    \Delta_{\sigma| \rho} \big(a \otimes I\big)|\rho ^{1/2} \rangle = \big(\mathcal{D}_{\sigma| \rho } (a)\otimes I \big)|\rho ^{1/2} \rangle.
\end{equation}
One can find $ \Delta_{\sigma| \rho} = \sigma \otimes \rho ^{-1}$ using the mirror operator. In case $ \sigma$ is the same as $ \rho$, the operator
\begin{equation}\label{34}
    \Delta_{\rho} = \rho \otimes \rho ^{-1} 
\end{equation}
corresponding to 
$\mathcal{D}_{\rho } (a) = \rho a \rho ^{-1}$ is called $modular$ operator.
It leaves $ |\rho ^{1/2} \rangle $ invariant. 
One can check that 
\begin{equation}\label{24}
    \begin{split}
          \Delta _\rho =& S _\rho S_\rho ^ \dagger\\
        J_\rho = &\Delta _\rho ^ {1/2} S_\rho \\
         S _\rho = J_\rho \Delta _\rho ^ {1/2} &= \Delta _\rho ^ {-1/2} J_\rho.
    \end{split}
\end{equation}
One can also show that     
    \begin{equation}
        J _\rho ~\mathcal{A}~ J_\rho = \mathcal{A}' ~~~~~~~~~~~~~~~~~~J _\rho ~\mathcal{A}' ~J_\rho = \mathcal{A}
    \end{equation}
and 
\begin{equation}
    ~~~\Delta ^ z _\rho ~\mathcal{A}~ \Delta ^ {-z}_\rho = \mathcal{A} 
         ~~~~~~~~~~~~~~~ \Delta ^ z _\rho ~\mathcal{A}' ~\Delta ^ {-z}_\rho = \mathcal{A}',
\end{equation}
for all $ z \in \mathbb{C}$. If we write $ \Delta _\rho = e ^ {-K_\rho}$, where $ K_\rho$ is called the  \emph{modular Hamiltonian}, the later equation can be interpreted as  
\begin{equation}
  e ^ {iK_\rho t } \mathcal{A}  e ^ {-iK_\rho t }= \mathcal{A} ~~~~~~~~~~~~
    e ^ {iK_\rho t } \mathcal{A}'  e ^ {-iK_\rho t } = \mathcal{A}'
\end{equation}
for $ z = -it$
which says both $\mathcal{A}$ and $ \mathcal{A}'$ are closed under time evolution using the modular Hamiltonian.

It is good to note it now that for every isometry $ v' \in \mathcal{A}'$,
the vector $ v'|\rho ^{1/2} \rangle$ or $ v'_m |\rho ^{1/2} \rangle$ is also a purification of $\rho$ in $ \mathcal{H}_\rho$. These vectors are also cyclic and separating for the algebra $ \mathcal{A}$. Thus, one could start from one of them instead of $|\rho ^{1/2} \rangle$ and build $ \mathcal{H}_\rho$ by acting the elements of $ \mathcal{A}$ on it. 
Actually, it comes from the point that while the eigenbasis of $ \rho$ is the preferred basis for $ \mathcal{K}$, we still have the freedom to choose a basis for $ \mathcal{K}'$. Here, acting with the isometry $ v'$ is indeed related to the change of basis in $ \mathcal{K}'$. 
To have just one unique vector corresponding to any state $\rho$, we can use the modular conjugation operator defined in (\ref{1}) or (\ref{2}), fix this operator, and choose the vector which is invariant under $ J_\rho$. The set of all vectors that are invariant under $ J_\rho$ is called the $natural$ $cone$. The states on $ \mathcal{A}$ are in one-to-one correspondence with the vectors in the natural cone. 
Take $ \ket{e}$ to be the vector corresponding to the maximally mixed state in the natural cone. For every $ \sigma \in \mathcal{S}(\mathcal{A})$, the vector 
$ (\sigma ^ {1/2} \otimes I) \ket{e}  $ is also in the natural cone:
\begin{equation}
    J_\rho ( \sigma ^ {1/2} \otimes I) \ket{e} = (I \otimes (\sigma ^ {1/2})^T)\ket{e} = ( \sigma ^ {1/2} \otimes I) \ket{e}.
\end{equation}
The vector $ \ket{e}$ itself is given as $ \ket{e}= ( \rho ^ {-1/2} \otimes I) |\rho ^{1/2} \rangle$ in $ \mathcal{H}_\rho$. Thus, the unique purification of the state $ \sigma$ in the natural cone is 
\begin{equation}
   |\sigma ^{1/2} \rangle = \big(\sigma ^ {1/2} \rho ^ {-1/2} \otimes I\big) |\rho ^{1/2} \rangle= \Delta _{\sigma | \rho} ^ {1/2} |\rho ^{1/2} \rangle.
\end{equation}
Thus, we reach to $ J_\rho |\sigma ^{1/2} \rangle = \Delta _{\sigma | \rho} ^ {1/2} |\rho ^{1/2} \rangle$, that also holds in infinite dimensional system. 

We usually consider the von Neumann algebra in its standard form which is defined as $ (\mathcal{A}, \mathcal{H}, J, \mathcal{P}_{\mathcal{A}} )$ where the algebra $ \mathcal{A}$ acts on the Hilbert space $\mathcal{H}$, $J$ is a anti-linear, unitary involution and $ \mathcal{P}_{\mathcal{A}}$ is the natural cone which is invariant by $ J$.

Finally, we note that although in a finite-dimensional system the Hilbert space approach and algebraic approach are equivalent, in an infinite dimension like QFT, it is not the case. In QFT, there is even no tensor factorization of the Hilbert space and Indeed the algebraic approach is appropriate to work in. 
For an open region, $  \mathcal{O}$ in the Minkowski spacetime, $\mathcal{A}_\mathcal{O} $ is defined to be the algebra of operators supported only in $\mathcal{O}$ which is called the \emph{local algebra} of the quantum field theory. $ \mathcal{A}_\mathcal{O}$ is also a von Neumann algebra that has the properties below:
\begin{enumerate}
    \item For $ \mathcal{O}_1 \subset \mathcal{O}_2$, we have $ \mathcal{A}_{\mathcal{O}_1} \subset \mathcal{A}_{\mathcal{O}_2} $.
    \item If $ \mathcal{O}_1 $ and $\mathcal{O}_2$ are spacelike seperated, we have  $ [\mathcal{A}_{\mathcal{O}_1}, \mathcal{A}_{\mathcal{O}_2}] = 0$.
    \item If $ \mathcal{O}' $ denote the causal complement of $ \mathcal{O}$, then $ \mathcal{A}_{\mathcal{O}}' = \mathcal{A}_{\mathcal{O}'}$, that is called \emph{Hagg duality}.
    \item If we denote the causal completion of $ \mathcal{O}$ as $ \Tilde{\mathcal{O}}$, then we have $ \mathcal{A}_{\Tilde{\mathcal{O}}} = \mathcal{A}_{\mathcal{O}}$.
\end{enumerate}
An important statement for local algebra in QFT is the \emph{ Reeh-Schlieder} theorem. It says that the vacuum vector $ \ket{\Omega}$ is cyclic and separating for the local algebra in any region $ \mathcal{O}$. It means that to generate the full vacuum sector of the Hilbert space, one needs to act just with the operator restricted to any arbitrary open region. Therefore, although there is not any notion of trace or tensor factorization in QFT, the Tomita-Takesaki theory provides us with a powerful tool to define the quantum information quantities also in QFT. 

As we had in Sec. \ref{Sec.Petz}, an important quantity to study the recoverability of the quantum channels in the theory of quantum error correction is  the relative entropy which is defined in \eqref{26}. But since, the expression in \eqref{26} can be used just for the Type \RNum{1} von Neumann algebra, to use the theory of QEC to study of QFTs and gravity, we need to generalaize the definition of relative entropy such that it can be applied for a generic type of von Neumann algebra.
One can check that
the relative entropy
can be also rewritten in terms of the relative modular operator in the GNS Hilbert space as 
\begin{equation}\label{25}
     S(\rho | \sigma) = -\langle \rho^ {1/2}| \log \Delta _{\sigma|\rho}|\rho^{1/2}\rangle.
\end{equation}
By using the expression \eqref{25},
the relative entropy was generalized to the general v.
Neumann algebras by Araki 
\cite{araki1975inequalities,araki1976relative}
using relative modular hamiltonians.
And in the case of the local algebra in QFTs,
the suitable definition of the relative entropy 
between two states $ \ket{\Psi}$ and $ \ket{\Phi}$ for measurements in the spacetime region $ \mathcal{O}$ is define as 
\begin{equation}
     S_\mathcal{O}(\Psi | \Phi) = -\langle \Psi| \log \Delta _{\Phi|\Psi}(\mathcal{O})|\Psi\rangle.
\end{equation}


\nocite{*}

\bibliographystyle{ieeetr}
\bibliography{refs}

\end{document}